\documentclass[usenatbib]{mnras}
\usepackage{adjustbox}
\usepackage{amsmath} 
\usepackage{amssymb} 
\usepackage{graphics}
\usepackage{graphicx}
\usepackage{grffile}
\usepackage{epsfig}
\usepackage{epstopdf}
\usepackage{times}
\usepackage[T1]{fontenc} 
\usepackage{aecompl}
\usepackage{grffile}
\usepackage{subfig}
\usepackage{multirow}
\usepackage{array}
\usepackage{tikz}
\usepackage{times}
\usepackage{tabularx}
\usepackage{adjustbox}
\usepackage{blindtext}
\usetikzlibrary{positioning}
\usepackage{marvosym}
\usepackage{float}
\usepackage{pdflscape}
\usepackage{afterpage}
\usepackage{rotating}
\usepackage{cite}
\usepackage{color}
\usepackage{longtable}

\newcommand{\todo}{\ifmmode \text{\Huge{\(\bullet\)}} \else {\Huge$\bullet$}\fi}
\newcommand{\tido}{\ifmmode {\bullet} \else $\bullet$\fi}

\newcommand{\ergs}	{\ifmmode {\text{erg\,s}}^{-1} \else erg s$^{-1}$\fi}
\newcommand{\kms}	{\ifmmode {\text{km\,s} }^{-1} \else km\,s$^{-1}$\fi}

\newcommand{\LCDM}{$\Lambda$CDM~}
\newcommand{\beq}{\begin{eqnarray}}  
\newcommand{\eeq}{\end{eqnarray}}

\newcommand{\ly}{{\ifmmode{{\text{Ly} }\alpha}\else{Ly$\alpha$}\fi}}

\newcommand{\hMpc  }{{\ifmmode{h^{-1}{\text{Mpc} }}\else{$h^{-1}$Mpc }\fi}}  
\newcommand{\hGpc  }{{\ifmmode{h^{-1}{\text{Gpc} }}\else{$h^{-1}$Gpc }\fi}}  
\newcommand{\hmpc  }{{\ifmmode{h^{-1}{\text{Mpc}  }}\else{$h^{-1}$Mpc }\fi}}  
\newcommand{\hkpc  }{{\ifmmode{h^{-1}{\text{kpc}  }}\else{$h^{-1}$kpc }\fi}}  
\newcommand{\hMsun }{{\ifmmode{h^{-1}{\text{M_{\rm \odot}}}}\else{$h^{-1}{\text{M_{\rm \odot}}}$}\fi}}   
\newcommand{\hmsun }{{\ifmmode{h^{-1}{\text{M_{\odot}}}}\else{$h^{-1}{\text{M_{\odot}}}$}\fi}}   
\newcommand{\Msun  }{\ifmmode M_{\rm \odot} \else $M_{\rm \odot}$\fi}  
\newcommand{\msun  }{\ifmmode M_{\rm \odot} \else $M_{\rm \odot}$\fi}

\newcommand{\rand  }{\ifmmode{{\mathcal{R}}}\else{${\mathcal{R}}$}\fi}

\newcommand{\Mbh   }{\ifmmode M_{\text{BH} } \else $M_{\text{BH} }$\fi}
\newcommand{\mbh   }{\ifmmode M_{\text{BH} } \else $M_{\text{BH} }$\fi}
\newcommand{\Mdot  }{\ifmmode \dot{M} \else $\dot{M}$\fi}
\newcommand{\LLedd }{\ifmmode L/L_{\text{Edd} } \else $L/L_{\text{Edd} }$\fi}
\newcommand{\lledd }{\ifmmode L/L_{\text{Edd} } \else $L/L_{\text{Edd} }$\fi}

\newcommand{\astar }{\ifmmode a_{*} \else  $a_{*}$\fi}
\newcommand{\RBLR  }{\ifmmode R_{\text{BLR} } \else $R_{\text{BLR}}$\fi}
\newcommand\abs[1]{\left|#1\right|}

\newcommand*{\rom}[1]{\expandafter\@slowromancap\romannumeral #1@}
\def\spose#1{\hbox to 0pt{#1\hss}}
\def\simlt{\mathrel{\spose{\lower 3pt\hbox{$\mathchar"218$}}
     \raise 2.0pt\hbox{$\mathchar"13C$}}}
\def\simgt{\mathrel{\spose{\lower 3pt\hbox{$\mathchar"218$}}
     \raise 2.0pt\hbox{$\mathchar"13E$}}}

\newcommand{  \Halpha   }{\ifmmode {\text{H} }\alpha \else H$\alpha$\fi}
\newcommand{  \ha   	}{\ifmmode {\text{H}}\alpha \else H$\alpha$\fi}
\newcommand{  \Hbeta    }{\ifmmode {\text{H} }\beta \else H$\beta$\fi}
\newcommand{  \hb    	}{\ifmmode {\text{H} }\beta \else H$\beta$\fi}
\newcommand{  \mgii     }{\ifmmode {\text{Mg} }\,\textsc{ii} \else Mg\,\textsc{ii}\fi}
\newcommand{  \MgII    }{\ifmmode {\text{Mg} }\,\textsc{ii}\,\lambda2798 \else Mg\,\textsc{ii}\,$\lambda2798$\fi}
\newcommand{  \HeII    }{\ifmmode {\text{He} }\,\textsc{ii}\,\lambda1640 \else He\,\textsc{ii}\,$\lambda1640$\fi}
\newcommand{  \NIII     }{\ifmmode {\text{N} }\,\textsc{iii}]\,\lambda1750 \else N\,\textsc{iii}]\,$\lambda1750$\fi}
\newcommand{  \NIV     }{\ifmmode {\text{N} }\,\textsc{iv}\,\lambda1718 \else N\,\textsc{iv}\,$\lambda1718$\fi}
\newcommand{  \NIVo     }{\ifmmode {\text{N} }\,\textsc{iv}]\,\lambda1486 \else N\,\textsc{iv}]\,$\lambda1486$\fi}
\newcommand{  \OIII     }{\ifmmode {\text{O} }\,\textsc{iii}]\,\lambda1663 \else O\,\textsc{iii}]\,$\lambda1663$\fi}
\newcommand{  \oiii     }{\ifmmode {\text{O} }\,\textsc{iii}]\,\lambda5007 \else O\,\textsc{iii}]\,$\lambda5007$\fi}
\newcommand{  \civ      }{\ifmmode {\text{C} }\,\textsc{iv}  \else C\,\textsc{iv}\fi}
\newcommand{  \ciii      }{\ifmmode {\text{C} }\,\textsc{iii}]  \else C\,\textsc{iii}]\fi}
\newcommand{  \CIV      }{\ifmmode {\text{C} }\,\textsc{iv}\,\lambda1549 \else C\,\textsc{iv}\,$\lambda1549$\fi}
\newcommand{  \CIII      }{\ifmmode {\text{C} }\,\textsc{iii}]\,\lambda1909 \else C\,\textsc{iii}]\,$\lambda1909$\fi}
\newcommand{  \SiOIV      }{\ifmmode {\text{Si} }\,\textsc{IV}+{\text{O} }\,\textsc{IV}]\,\lambda1400 \else Si\,\textsc{iv}+O\,\textsc{iv}]\,$\lambda1400$\fi}
\newcommand{  \sioiv      }{\ifmmode {\text{Si} }\,\textsc{IV}+{\text{O} }\,\textsc{IV}]\ \else Si\,\textsc{iv}+O\,\textsc{iv}]\fi}
\newcommand{  \feii      }{\ifmmode {\text{Fe} }\,\textsc{ii}  \else Fe\,\textsc{ii}\fi}

\newcommand{  \FeII      }{\ifmmode {\text{Fe} }\,\textsc{ii}  \else Fe\,\textsc{ii}\fi}

\newcommand{  \FeIII      }{\ifmmode {\text{Fe} }\,\textsc{iii}  \else Fe\,\textsc{iii}\fi}

\newcommand{ \Lha   }{\ifmmode L\left(\ha\right) \else $L\left(\ha\right)$\fi}
\newcommand{ \fwha  }{\ifmmode {\text{FWHM} }\left(\ha\right) \else FWHM(\ha)\fi}

\newcommand{ \Lmg   }{\ifmmode L\left(\mgii\right) \else $L\left(\mgii\right)$\fi}
\newcommand{ \fwmg  }{\ifmmode {\text{FWHM} }\left(\mgii\right) \else FWHM(\mgii)\fi}
\newcommand{ \Lciv  }{\ifmmode L\left(\civ\right) \else $L\left(\civ\right)$\fi}
\newcommand{ \fwciv }{\ifmmode {\text{FWHM} }\left(\civ\right) \else FWHM(\civ)\fi}
\newcommand{ \fwhm  }{\ifmmode {\text{FWHM} } \else \text{FWHM}\fi} 
\newcommand{ \voff  }{\ifmmode v_{\text{off} } \else $v_{\text{off} }$\fi} 

\newcommand{ \fwhb  }{\ifmmode {\text{FWHM} }\left(\hb\right) \else FWHM(\hb)\fi}

\newcommand{ \sigline  }{\ifmmode \sigma_{\text{line}} \else $\sigma_{\text{line}}$\fi}

\newcommand{ \sigmamg  }{\ifmmode {\sigma }\left(\mgii\right) \else $\sigma$(\mgii)\fi}
\newcommand{ \sigmaciv  }{\ifmmode {\sigma }\left(\civ\right) \else $\sigma$(\civ)\fi}
\newcommand{ \sigmahb  }{\ifmmode {\sigma }\left(\Hbeta\right) \else $\sigma$(\Hbeta)\fi}
\newcommand{ \sigmaha  }{\ifmmode {\sigma }\left(\Halpha\right) \else $\sigma$(\Halpha)\fi}
\newcommand{ \sigmas  }{\ifmmode {\sigma }_{\star} \else $\sigma_{\star}$\fi}

\newcommand{\MbhHb   }{\ifmmode M_{\text{BH}  } \left( \Hbeta \right) \else $M_{\text{BH} } \left( \Hbeta \right)$\fi}
\newcommand{\MbhHa   }{\ifmmode M_{\text{BH} }  \left( \Halpha \right) \else $M_{\text{BH} } \left( \Halpha \right)$\fi}%
\newcommand{\MbhMg   }{\ifmmode M_{\text{BH}  } \left( \mgii \right) \else $M_{\text{BH} } \left( \mgii \right)$\fi}
\newcommand{\MbhC   }{\ifmmode M_{\text{BH}  } \left( \civ \right) \else $M_{\text{BH} } \left( \civ \right)$\fi}

\newcommand{\Mbhfw   }{\ifmmode M_{\text{BH}}\left(\text{FWHM}\right)    \else $M_{\text{BH}}\left(\text{FWHM}\right)$\fi}
\newcommand{\Mbhsig   }{\ifmmode M_{\text{BH}}\left(\sigma_{\text{line}}\right)    \else $M_{\text{BH}}\left(\sigma_{\text{line}}\right)$\fi}

\newcommand{ \fwcm  }{\ifmmode \fwhm\left[\civ/\mgii\right] \else $\fwhm\left[\civ/\mgii\right]$\fi}
\newcommand{ \FWCM  }{\ifmmode \fwciv/\fwmg \else $\fwciv/\fwmg$\fi}

\newcommand{ \fwch  }{\ifmmode \fwciv/\fwhb \else $\fwciv/\fwhb$\fi}
\newcommand{ \fwcha  }{\ifmmode \fwhm\left[\civ/\Halpha\right] \else $\fwhm\left[\civ/\Halpha\right]$\fi}

\newcommand{ \dvmax }{\ifmmode \Delta v_{\rm peak}\left(\civ\right) \else $\Delta v_{\rm peak}\left(\civ\right)$\fi}

\newcommand{ \dvline }{\ifmmode \Delta v_{\rm line}\left(\civ\right) \else $\Delta v_{\rm line}\left(\civ\right)$\fi}

\newcommand{ \dvh }{\ifmmode \Delta v_{\rm 50}\left(\civ\right) \else $\Delta v_{\rm 50}\left(\civ\right)$\fi}

\newcommand{ \dvt }{\ifmmode \Delta v_{\rm 90}\left(\civ\right) \else $\Delta v_{\rm 90}\left(\civ\right)$\fi}

\newcommand{\local}{ \textit{local\ } }

\newcommand{\fw}{\ifmmode \fwhm_{\text{local}} \else $\fwhm_{\text{local}}$\fi}
\newcommand{\fwtdc   }{\ifmmode \fwhm_{\text{global}} \else $\fwhm_{\text{global}}$\fi}
\newcommand{\Llocal   }{\ifmmode L_{\text{local}} \else $L_{\text{local}}$\fi}
\newcommand{\Ltdc   }{\ifmmode L_{\text{global}} \else $L_{\text{global}}$\fi}

\newcommand{ \mumg  }{\ifmmode \mu\left(\mgii\right) \else $\mu\left(\mgii\right)$\fi}
\newcommand{ \fmg   }{\ifmmode f\left(\mgii\right) \else $f\left(\mgii\right)$\fi}
\newcommand{ \muciv }{\ifmmode \mu\left(\civ\right) \else $\mu\left(\civ\right)$\fi}
\newcommand{ \fciv  }{\ifmmode f\left(\civ\right) \else $f\left(\civ\right)$\fi}

\newcommand{  \Luv      }{\ifmmode L_{1450} \else $L_{1450}$\fi}
\newcommand{  \Lop      }{\ifmmode L_{5100} \else $L_{5100}$\fi}
\newcommand{  \Loploc      }{\ifmmode L_{5100}^{\text{local}} \else $L_{5100}^{\text{local}}$\fi}
\newcommand{  \Lopglob      }{\ifmmode L_{5100}^{\text{global}} \else $L_{5100}^{\text{global}}$\fi}
\newcommand{  \Lsix      }{\ifmmode L_{6200} \else $L_{6200}$\fi}
\newcommand{  \Lthree   }{\ifmmode L_{3000} \else $L_{3000}$\fi}

\newcommand{  \lpcs      }{\ifmmode L_{\rm peak}\left(\civ \right)/L_{\rm peak}\left(\sioiv \right) \else $ L_{\rm peak}\left(\civ \right)/L_{\rm peak}\left(\sioiv \right)$\fi}

\newcommand{  \lpcc      }{\ifmmode L_{\rm peak}\left(\civ \right)/L_{\rm peak}\left(\ciii \right) \else $ L_{\rm peak}\left(\civ \right)/L_{\rm peak}\left(\ciii \right)$\fi}

\newcommand{  \lpc      }{\ifmmode L_{\rm peak}\left(\civ \right)/L\left(1450\text{\AA} \right) \else $L_{\rm peak}\left(\civ \right)/L\left(1450\text{\AA} \right)$\fi}

\newcommand{  \LPcs      }{\ifmmode  L_{\rm peak}\left[\civ/{\rm SiOIV} \right] \else $L_{\rm peak}\left[\civ/{\rm SiOIV} \right]$\fi}
\newcommand{  \LPcc      }{\ifmmode  L_{\rm peak}\left[\civ/\ciii \right] \else $L_{\rm peak}\left[\civ/\ciii \right]$\fi}
\newcommand{  \LPc      }{\ifmmode  L_{\rm peak}\left[\civ/1450\text{\AA} \right] \else $L_{\rm peak}\left[\civ/1450\text{\AA} \right]$\fi}

\hypersetup{draft}
\begin{document}
\title[\civ-based Black Hole Mass estimation]
{ Can we improve \civ-based single epoch black hole mass estimations?}

\author[ J. E. Mej\'ia-Restrepo et al]{J. E. Mej\'ia-Restrepo,$^{1,2}$\thanks{Email: jmejiar@eso.org} B. Trakhtenbrot,$^{3}$ \thanks{Zwicky postdoctoral fellow} P. Lira$^{1}$ and H. Netzer$^{4}$  \vspace*{6pt}\\ $^{1}$ Departamento de Astronom\'{i}a, Universidad de Chile, Camino el Observatorio 1515, Santiago, Chile \\$^{2}$ European Southern Observatory, Casilla 19001, Santiago 19, Chile\\ $^{3}$ 
Department of Physics, ETH Zurich, Wolfgang-Pauli-Strasse 27, CH-8093 Zurich, Switzerland\\ $^{4}$ School of Physics and Astronomy, Tel Aviv University, Tel Aviv 69978, Israel} 

\maketitle

\begin{abstract}
 In large optical surveys	at high redshifts ($z>2$), the \CIV\ broad emission line is the most practical alternative to estimate the mass (\Mbh) of active super-massive black holes (SMBHs). However, mass determinations obtained with this line are known to be highly uncertain. In this work we use the Sloan Digital Sky Survey Data Release 7 and 12  quasar catalogues to statistically test three  alternative methods put forward in the literature to improve  \civ-based \Mbh\ estimations. These methods are constructed from correlations between the ratio of the \CIV\ line-width to the low ionization line-widths (\Halpha, \Hbeta\ and \MgII) and several other properties of rest-frame UV emission lines.  Our analysis suggests  that  these correction methods are of limited applicability, mostly because all of them depend on correlations that are driven by the linewidth of the \civ\ profile itself  and not by an interconnection between the linewidth of the \civ\ line with the linewidth of the low ionization lines.
Our results show that optical \civ-based mass estimates at high redshift cannot be a proper replacement for estimates based on IR spectroscopy of low ionization lines like \Halpha, \Hbeta\ and \mgii.

\end{abstract}
 

\section{Introduction}
\label{sec:intro}

Accurate determinations of super-masssive black hole masses (\Mbh s) are essential to fully understand SMBH the physics, demographics, and relations with galaxies. 
 The single epoch (SE)  black hole mass estimation method is commonly used  on large samples of  unobscured, type-I active galactic nuclei  \citep[AGN;][]{MclureDunlop2004,Onken2008,Shen2008,Fine2010,RafieeHall2011,TrakhtenbrotNetzer2012}.
This method relies on two basic ingredients: (1) the assumption of virialized gas kinematics in the  broad line region (BLR) and  (2) the empirical relation from reverberation mapping (RM) experiments between the BLR size (\RBLR) and the continuum luminosity ($L_{\lambda}\equiv \lambda L\left( \lambda \right)$) at al particular wavelength ($\lambda$) where $\RBLR  \propto  (L_{\lambda} )^{\alpha} $ with $\alpha\sim0.5-0.7$ \citep{Kaspi2000,Kaspi2005,Bentz2009,Park2012,Bentz2013}.

Under these assumptions the width of the broad emission lines, such as the full width at half maximum (\fwhm), is a good proxy for the virial velocity of the BLR clouds. \Mbh\  can thus be 
 expressed as:
\begin{equation}
 \Mbh =f G^{ - 1} R_{\text{BLR} } v_{\text{BLR}}^{2} = K (L_{\lambda} )^{\alpha}\fwhm^{2}.
\end{equation}
Here $G$ is the gravitational constant, $f$ is a geometrical factor that  accounts for the unknown structure and inclination to the line 
of sight of the BLR.   In this paper,  we  assume $f=1$, which is an appropriate median value  for  \Mbh\ estimates using  the \fwhm\ \citep{Woo2015}. 
However, there  is a large uncertainty  in this value \citep[of at least a factor of 2; e.g.][]{Onken2004,Woo2013,Shankar2016,Batiste2017}   that can be  even larger if $f$ depends on  luminosity and/or  other line 
properties  \citep[e.g. equivalent widths, line offsets, \fwhm;][]{Collin2006,Shen2013,MejiaRestrepo2017}.

The most reliable RM-based $\RBLR-L$ relation is the $R_{\rm BLR}\left(\Hbeta\right)-\Lop$ relation. This relation is the only one that has been established for a  large number of sources and covering a broad luminosity range ($10^{43}\,\ergs\lesssim\Lop\lesssim 10^{46}\,\ergs$).  Consequently, SE  \Mbh\ calibrations for   other lines are often  re-calibrated to match \Mbh\ measurements based on \Hbeta\ and \Lop.  Such  re-calibrations are used to determine \Mbh\ values at different redshifts where the \Hbeta\ is not available due to observational limitations. In optical surveys, the   \Halpha\ and \Hbeta\ lines can be used up to $z\lesssim0.8$ \citep[e.g.][]{GreeneHo2005,NetzerTrakhtenbrot2007,Xiao2011,ShenLiu2012},  the \MgII\ (hereafter \mgii) can assist for \Mbh\ on sources where $0.6 \lesssim z\lesssim2.2$  \citep[e.g.,][]{MclureJarvis2002,VestergaardOsmer2009,Wang2009,Trakhtenbrot2011,ShenLiu2012,TrakhtenbrotNetzer2012} and the \CIV\ line (hereafter \civ) is used to estimate black hole masses at even higher redshifts   \citep[$2.0\lesssim z\lesssim5.0$;][]{VestergaardPeterson2006,Park2013}.

\Mbh\ calibrations based on low ionization lines (i.e. \Halpha,  and \mgii) generally show good agreement with the \Hbeta\ \Mbh\ estimator with a typical scatter of $\lesssim 0.2$ dex \citep{GreeneHo2005,Xiao2011,TrakhtenbrotNetzer2012}. However, the  analogous recalibration using the  \civ\  high ionization line is  more problematic and shows large scatter (0.4-0.5 dex), possibly  driven by  several causes. First,  the width of \civ\ is only weakly correlated, if at all, with the width of  the low ionization lines  and presents large scatter in many AGN samples \citep[e.g.,][]{BaskinLaor2005,Netzer2007,Shang2007,Shen2008,Fine2010,Ho2012,ShenLiu2012,TrakhtenbrotNetzer2012,Tilton2013}.
Second, the \civ\  profiles show  large blue-shifts  \citep[up to several thousands \kms; ][]{Richards2002,BaskinLaor2005,Shang2007,Richards2011} that indicates  non virial motions. Third,  \citet{Denney2012}  found that the  core of the broad \civ\ line does not reverberate in response to continuum variations.  This implies that not only the innermost but also the \emph{outermost} \civ\ emitting regions may not be virialized.

Given that \civ\ is the most widely used line  for \Mbh\ determination at $z\gtrsim2$  in optical surveys, it is crucial to design practical methodologies to mitigate the issues related to \civ-based \Mbh\ determinations.  There have been  many efforts to improve single-epoch \Mbh\ determinations from \civ\ \citep[e.g.,][]{VestergaardPeterson2006, Assef2011,Denney2013, Park2013,Runnoe2013, Tilton2013, Brotherton2015, Coatman2016}.   
The studies of \citet{Assef2011}, \citet{Denney2013}, \citet{Park2013}, and \citet{Tilton2013} claimed that in spectra of limited signal to noise (S/N) and/or spectral resolution, \fwciv\ measurements are underestimating the ``real'' line widths, in objects with strong intrinsic absorption features that cannot be deblended from the emission lines. 
This would partially explain the fact that about 40\% of  quasars show  \civ\ profiles narrower than the \Hbeta\ profiles   \citep{TrakhtenbrotNetzer2012}, in contrast to expectations from RM experiments which finds that the bulk of the \civ\ emitting region is interior to that of the low ionization lines \citep[][Lira et al. 2017 submitted]{Peterson2005,Kaspi2007,Trevese2014}.  However, \citet{CorbinBoroson1996} found that even objects with no evidence of absorption features show \fwciv$<$\fwhb.

 Additionally, \citet{Assef2011}, \citet{Denney2013}, \citet{Park2013} and \citet{Tilton2013} 
also found that, in high quality spectrum (S/N$\gtrsim 10$, R$\gtrsim 2000$), the line dispersion (\sigline) of the \civ\ line
 provides \civ-based \Mbh\ estimations in better agreement with \Hbeta-based estimations than the \fwhm\ of the \civ\ line. However, in \citet{MejiaRestrepo2016a} we analysed the impact of using different approaches to trace the continuum emission underneath the emission lines in estimating \sigline\ and \fwhm\ of broad emission lines. Our results indicate that \sigline\  measurements are very sensitive by the continuum determination even in high quality spectra. On the other hand, we found that \fwhm\ measurements are only weakly  affected by this effect ever since the line profile is no affected by strong absorption feature. This result suggests the usage of the \fwhm, instead of \sigline, for \civ-based \Mbh\ estimations after neglecting those objects with strong absorption features.

Recently, \citet{Coatman2017} found a strong correlation  between the blue-shift of the \civ\ line centroid and the \civ/\Halpha\ FWHM ratio for a sample of  66 high luminosity ($10^{46.5}\ \ergs<L_{\rm Bol}<10^{47.5}\ \ergs$) and high redshift quasars ($z>2.1$). They suggested that this correlation can assist to improve \civ-based black-hole masses reducing the scatter between \civ\ and \Halpha\ based \Mbh\ determinations from  0.40 dex to  0.24 dex. However, this procedure is not applicable to large optical surveys because of the difficulty to accurately determine the AGN redshift, necessary to compute the \civ\ blueshift, without information from low ionization lines.  \citet{Runnoe2013} and \citet{Brotherton2015} used a sample of 85  low-redshift ($0.03<z<1.4$) and low-to-moderate luminosity ($10^{43.4}\ \ergs<\Lop<10^{46.5}\ \ergs$) AGN with quasi-simultaneous UV and optical rest-frame spectra to propose a method to rehabilitate \civ\ for \Mbh\ determination. The method consisted of using a correlation that they found between the  \SiOIV$/$\civ\  line peak intensity ratio and the \Hbeta$/$\civ\ \fwhm\ ratio.  They claim that using this correlation it is possible to predict \fwhb\ from measurements of the \SiOIV\ (hereafter \sioiv) emission to obtain more accurate \civ\ based mass measurements. They specifically claim that the scatter between \civ\ and \Hbeta\ estimations is reduced  from 0.43 dex to 0.33 dex.

In our previous studies, we presented a  sample of 39  high-quality, simultaneous (rest-frame) UV-optical spectra of  type-1 AGN at z$\sim$1.5  obtained with X-Shooter \citep{Capellupo2015,Capellupo2016}. Using this sample, in  \citet{MejiaRestrepo2016a} we  were able to reproduce  the  correlation found in \citet{Runnoe2013}  but with a weaker statistical significance.  \citet{MejiaRestrepo2016a}  also found a similar but  alternative correlation between the   \CIII/\civ\  line peak intensity ratio and the \Hbeta$/$\civ\ \fwhm\ ratio.  In general, we found that  the ratios of \fwciv\ to the \fwhm\ of the  \Halpha, \Hbeta\ and \mgii\  low ionization lines are correlated with 
both, the \sioiv$/$\civ\ and the  \CIII$/$\civ\ line peak ratios. In spite of these correlations, we found that none of them are able to reduce the scatter between \civ-based \Mbh\ estimations and the low ionization line  \Mbh\ estimations. 

 It is important to point out that the findings of  \citet{Coatman2017}, \citet{MejiaRestrepo2016a} and \citet{Runnoe2013}   are all obtained from relatively small samples (230, 69 and 39 objects respectively) that map different regions in the parameter space of the AGN population (see \S \ref{sec:data1}).  Thus, the  significance of their findings may  not be applicable to the overall population of  non-obscure type-I  AGN population. 
 
The purpose of this work is  to test  the validity of these empirical alternatives to improve \civ-based \Mbh\ estimations  on large AGN samples with survey-grade spectroscopic data. To accomplish this goal, in this paper we  use data from the Sloan Digital Sky Survey \citep[SDSS][]{York00}, specifically from the SDSS-II data release 7 and the SDSS-III data release 12 quasar spectroscopic catalogues  \citep[DR7Q and DR12Q respectively,][]{Schneider2010,Paris2017}. 
 All the alternatives that we are testing here stand on correlations that relate \fwch\ with  the \civ\  line itself or with  the properties of  emission lines or  continuum windows that are close to  the \civ\  line. 
Due to the lack of simultaneous coverage of \civ\ and \Hbeta\ lines in the optical SDSS survey, we will carry our analysis in terms of  $\FWCM\equiv\fwcm$ instead of \fwch. This is justified as it is well known that \fwmg\  is tightly correlated with \fwhb\  and that \Mbh\ estimations from these two emission lines are known to agree within 0.2 dex of accuracy \citep[e.g][]{Wang2009,ShenLiu2012,TrakhtenbrotNetzer2012,MejiaRestrepo2016a}. Because of the limitation in S/N of the SDSS data and the difficulties in measuring the \sigline\ of \civ,  in this paper we do not thoroughly explore the usage of this quantity for \civ\ measurements, we however briefly analyse the feasibility of its usage in a high-quality spectra subsample of the SDSS DR12Q catalogue.

This paper is structured as follows. In section \S\ref{sec:data1} we present the  samples  and introduce the most relevant parameters that we measured for our analysis.  In \S\ref{sec:results1} we present and discuss our main results and in \S\ref{sec:conclusions} we highlight our most important findings.  Throughout this paper we assume a flat \LCDM\ cosmology with the following values for the cosmological parameters: $\Omega_{\Lambda}= 0.7$, $\Omega_{\text{M} }= 0.3$ and $H_{0}=70\,\kms\,\rm{Mpc}^{-1}$.

\section{Samples, Data and Analysis}
\label{sec:data1}

\begin{figure*}
  \includegraphics[width=\columnwidth]{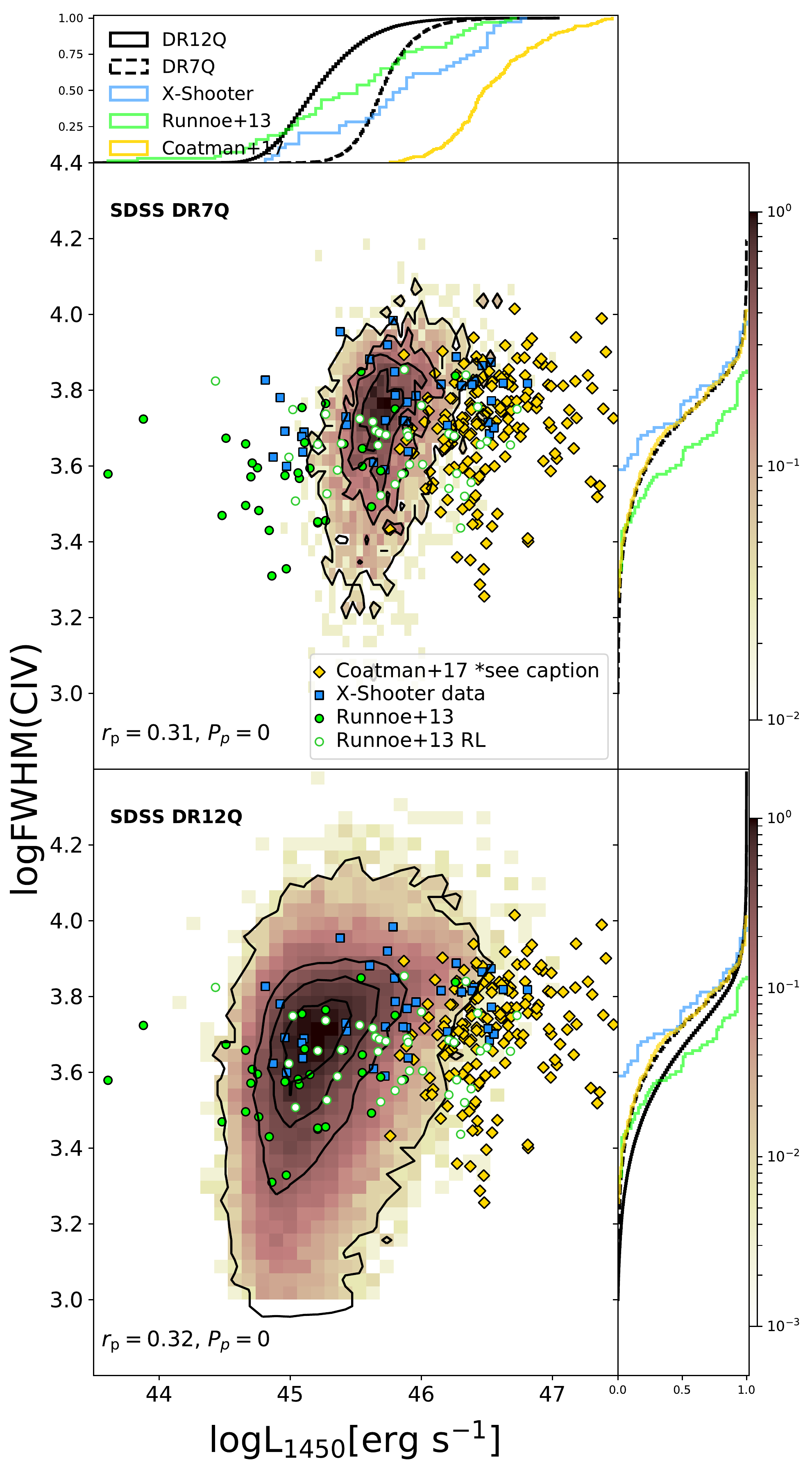}
 \includegraphics[width=\columnwidth]{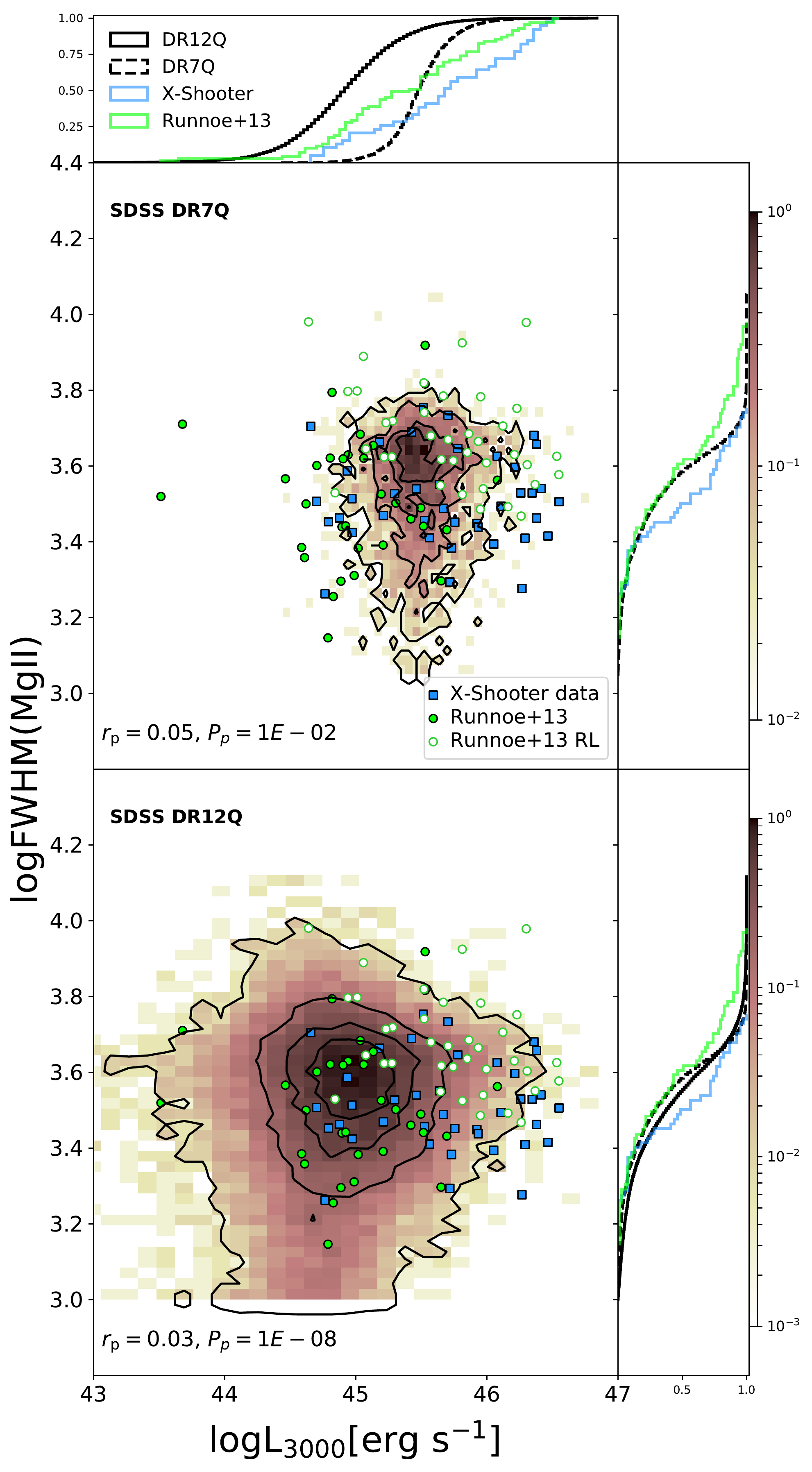} \\
  \caption{ Left column: \Luv-\fwciv\ bi-dimensional  distributions  in the SDSS DR7Q (top-left) and DR12Q  (bottom-left) samples. The intensity of the colour represents the relative density of points as shown in the colour bars on the right. The black thin lines represent the 
  25\%, 50\%, 75\% and 99\%   contours centred at the maximum probability point. In the top and right side diagrams we show the projected  CDFs of \Luv\  and \fwciv, respectively. We superimpose in each panel analogue data from  the small X-Shooter, R13 and C17 samples as summarized in the legend. R13 is differentiated by RL and RQ objects in the scatter plots. For the C17 sample  we show $L_{\rm 1350}$ as a proxy for $L_{\rm 1450}$  ($L_{\rm 1450}$ is not listed in \citet{Coatman2016}).  We also show the Pearson correlation coefficient (r$_{\rm P}$) and associated probability for upholding the null hypothesis (P$_{\rm P}$) for the  DR7Q and DR12Q bi-dimensional distribution.  Right column: \Lthree-\fwmg\ bi-dimensional  distributions. Description is identical to the left column. C17 data are not available for these quantities.     }
 \label{fig:pdfL1450}
 \end{figure*}
 
 In this section  we describe in detail two large samples, namely, the  SDSS DR7Q and the DR12Q samples,  as well as three  small samples taken from   \citet{MejiaRestrepo2016a} , \citet{Runnoe2013} and \citet{Coatman2017}. We also describe the spectral   fitting procedure and  the the emission line and continuum properties that were obtained for the analysis presented here.

\subsection{Large Samples}

To accomplish our  goal, we need to guarantee the simultaneous coverage of the \sioiv, \civ, \ciii, and \mgii\ emission lines. According to the spectroscopic coverage of the DR7Q  (3800-9200\AA) and the DR12Q (3600-10400\AA) samples, we selected objects with 1.8$<$z$<$2.0 and  $1.70<z<2.3$ respectively. These redshift constraints translate into a total of 4817 objects for the DR7Q catalogue  and 69092 objects for the DR12Q catalogue.  Although the objects from the SDSS DR7 sample are also included in the DR12Q 
sample, to construct the DR12Q catalogue, the objects from the DR7Q catalogue were re-observed. The time interval between observations is at least 4 years.  

For the DR7Q sample we used the redshifts estimations from \citet{HewettWild2010} which provides important improvements to the SDSS redshifts estimations with a reduction of a factor of 6 of the systematic uncertainties with respect to SDSS redshift estimations. For the DR12Q sample, \citet{HewettWild2010} redshift calculations are not available. We therefore adopt the  visually inspected  SDSS redshifts  estimations described  in \citet{Paris2017}. Consequently, line shift estimations in the SDSS DR12Q sample are less reliable than in the DR7Q sample. 

It is also important to note that because the survey was designed  to map the large scale structure of the universe at high redshift, the DR12Q catalogue is  primarily biased towards $z>2$ sources \citep{BOSS2009}. In particular, in our sub-sample of the DR12Q catalogue  75\% of the objects are at $z>2.0$.
 
\begin{figure*}
  \includegraphics[width=\linewidth]{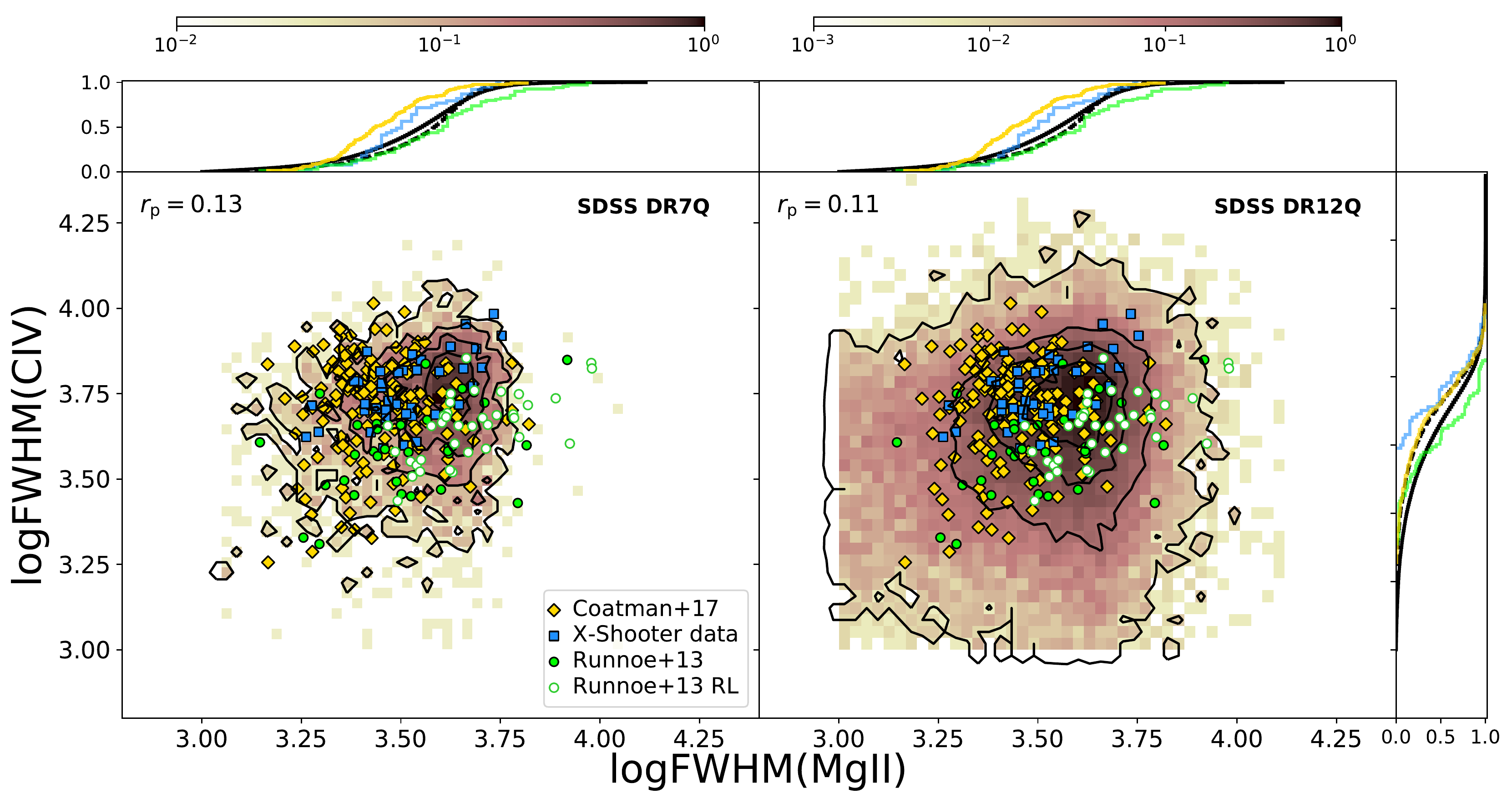}
  \caption{ \fwmg-\fwciv\ bi-dimensional  distributions  in the SDSS DR7Q (left) and DR12Q  (right) samples. The intensity of the colour represents the relative density of points as shown in the colour bars on the right. The black thin lines represent the 
  25\%, 50\%, 75\% and 99\%   contours centred at the maximum probability point. In the top and right side diagrams we show the projected  CDFs of \fwmg\  and \fwciv, respectively. We superimpose in each panel analogue data from  the small X-Shooter, R13 and C17 samples as in Figure \ref{fig:pdfL1450}. R13 is differentiated by RL and RQ objects in the scatter plots.  We also show the Pearson correlation coefficient (r$_{\rm P}$) for the  DR7Q and DR12Q bi-dimensional distribution.      }
 \label{fig:pdfFWCIVFWMG}
 \end{figure*}

\subsection{Small Samples}
We complement our analysis with three additional smaller samples with considerably higher-quality spectroscopic data.  These samples correspond to the original samples  used to propose the different methodologies  to improve \civ-based \Mbh\ estimations that we described in the introduction. 

The first sample is described in \citet[hereafter the X-Shooter sample]{Capellupo2016} consisting of 39 RQ quasars  observed by the X-Shooter spectrograph that guaranteed  simultaneous observations of the rest-frame UV and optical. The sample comprises objects with  $1.45<z<1.69$ and $10^{44.8}\ergs<\Luv<10^{46.8}\ergs$. 

The second of these samples is described in \citet[hereafter the R13 sample]{Runnoe2013} consisting of 69 objects  including 37 radio-loud (RL) and 32 radio-quiet (RQ) quasars with nearly simultaneous observations of the (rest-frame) X-ray, Ultraviolet (UV) and optical. This sample is a subset of the \citet{Tang2012} sample and comprises objects with $0.03<z<1.4$ and $10^{43.6}\ergs<\Luv<10^{46.7}\ergs$.

Finally, the sample described in \citet[ hereafter the C17 sample]{Coatman2017} consists of a compilation of  230 RQ quasars with $10^{45.7}\ergs<L_{\rm 1350}<10^{47.7}\ergs$. This sample comprises sources from \citet{ShenLiu2012},  \citet{Coatman2016} and \citet{Shen2016}. All the sources have non-simultaneous optical observations (from SDSS) and ground based near infra-red observations  which at the redshift range of the sample ($1.5<z<4.0$) correspond to the rest-frame UV and optical range, respectively. This sample has not reported \mgii\ emission line measurements but includes \Halpha\ emission line measurements that can be used as a proxy for \mgii\ line measurements \citep[see e.g.,][]{ShenLiu2012,MejiaRestrepo2016a}. 

\subsection{Line and continuum measurements}

For each object in the SDSS DR7Q and DR12Q samples we fitted the  line  profiles of the  \sioiv, \civ, \ciii\ and \mgii\   emission lines as  described in Appendix \ref{app:fit}. From the best fit model of the emission lines of each object  we measured the line  \fwhm, the velocity dispersion  \citep[\sigline; following][]{Peterson2004}, the rest-frame equivalent width ($EW\left({\rm line}\right)$), the integrated line luminosity ($L\left({\rm line}\right)$) and the luminosity at the peak of the fitted profile ($L_{\text{peak}}\left({\rm line}\right)\equiv 4\pi D_{\rm L}^2 F_{\rm \lambda, peak} ({\rm line})$). As line blue-shift indicators we measured two different quantities: (1) the shift of the emission line peak ($\Delta v_{\rm peak}$) and  (2) the line centroid shift defined as   shift in the flux-weighted central wavelength \citep[$\Delta v_{\rm line}$, following][]{Peterson2004}.
We also computed the monochromatic luminosities at different wavelengths ($L_{\lambda}\equiv \lambda\ L\left(\lambda \right)$). We particularly measured $L_{1350}$, \Luv, $L_{2000}$ and \Lthree\  that correspond to   continuum bands adjacent to the \sioiv , \civ, \ciii\ and the \mgii\ emission lines, respectively.  Finally, from the large DR7Q and DR12Q samples we excluded  broad absorption line quasars (BALQSOs) and objects with unreliable fits following the strategy described in Appendix \ref{app:fit}. We ended up with 3267 objects from the DR7Q catalogue (out of 4817) and 35674 from the DR12Q catalogue (out of 69062 objects).

In the case of the  X-Shooter, R13 and C17 samples  we also extracted  the measurements of the aforementioned quantities whenever available from the published data in \citet{MejiaRestrepo2016a}, \citet{Runnoe2013} and \citet{Coatman2017}, respectively.    Although the fitting approaches in each of these papers are not identical, they follow similar procedures  and then provide comparable measurements. 

\citet{MejiaRestrepo2016a}  showed that \sigline\ and $L\left({\rm line}\right)$ are very sensitive to the continuum placement because of their strong  dependence on the line wings. Analogously, \dvline, one of the most widely used blue-shift indicators, is also affected by the continuum placement. This fact motivates us to use the alternative blue-shift estimators $\Delta v_{\rm peak}$ 
(see definition above).  Similarly,  $EW\left({\rm line}\right)$ is also sensitive to continuum placement. Therefore, in addition  to $EW\left({\rm line}\right)$, we also  use $L_{\text{peak}}\left({\rm line}\right)/L_{\lambda} $ because of its  weaker dependency on continuum placement.

We thus have a  set of quantities that are weakly sensitive to continuum placement given by \fwhm, $L_{\text{peak}}\left(\text{line}\right)$, $\Delta v_{\text{peak}}$ and $L_{\text{peak}}\left({\rm line}\right)/L_{\lambda} $, as well as a set of quantities that are strongly affected by the placement of the continuum emission given by \sigline, $L\left({\rm line}\right)$, $\Delta v_{\rm line}$ and $EW\left({\rm line}\right)$. We emphasize however that the latter quantities are also important because they carry information about the broadest components of the emission lines and therefore  of the innermost region of the BLR. 

\begin{figure*}
      \includegraphics[width=1.0\textwidth ,angle =0,keepaspectratio]{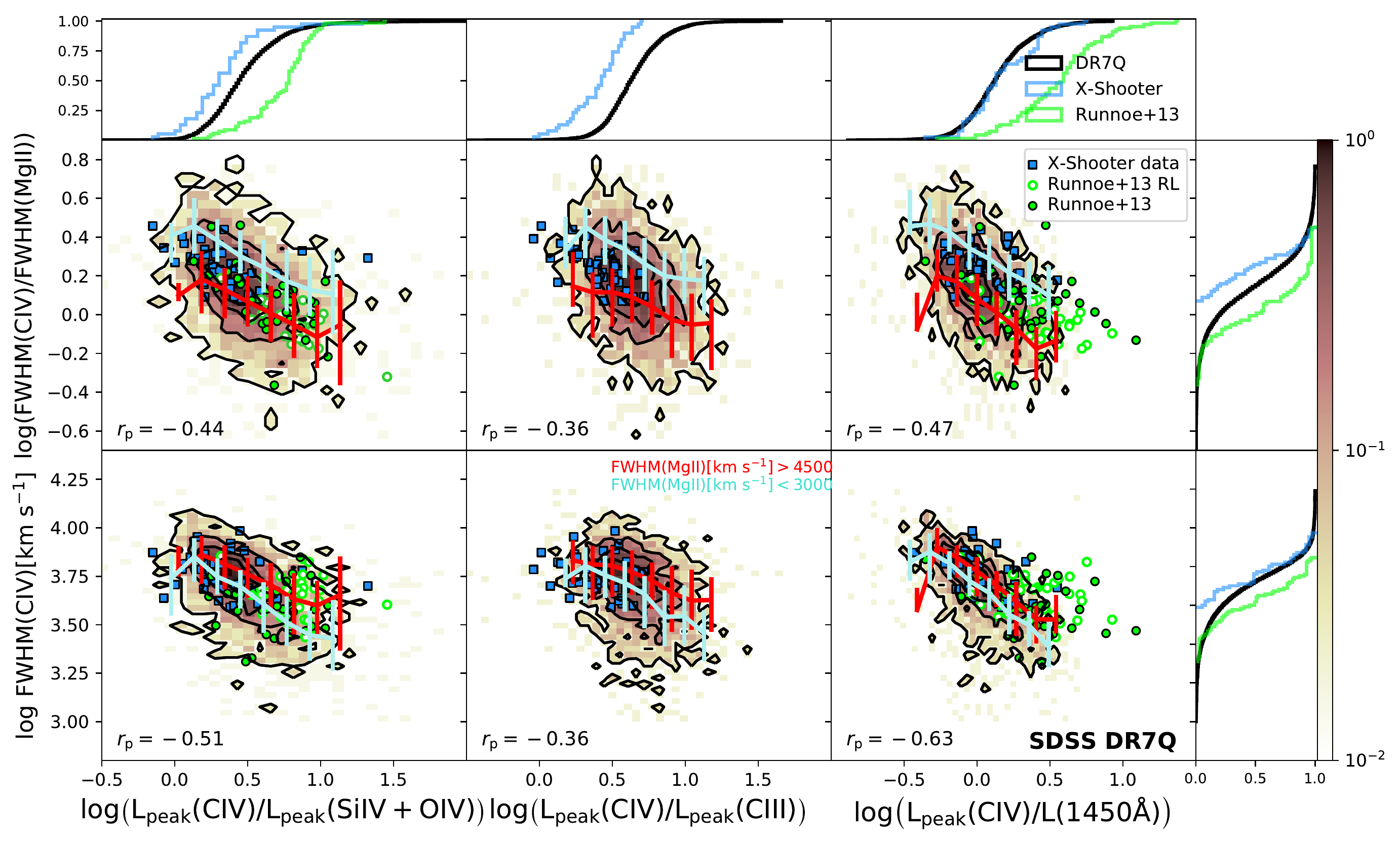}
 
 \caption{ Bidimensional distributions of \fwcm\ (top-row) and  \fwciv\ (bottom row) vs 
 \LPcs\ (left-column), \LPcc\ (centre-column) and \LPc\ (right-column)
 in the {\bf SDSS DR7Q} sample. The intensity of the colour represents the relative density of points as shown in the colour bar on the right. The black thin lines represent the   25\%, 50\%, 75\% and 99\%   contours centred at the maximum probability point. The projected CDFs  of each of the quantities are also shown in the right  and top side diagrams. We superimpose to each panel analogue data of the  X-Shooter and R13  small samples   as indicated in the legends. Coloured trend lines represent the median values of  \fwcm\ (middle panels) and \fwciv\ (bottom panels) as a function of the different line peak ratios  for  objects  with $\fwmg<3000\ \kms$ (light-turquoise) and $\fwmg>4500\ \kms$ (red). The error bars represent the 1-$\sigma$ dispersion of the points around these trends. Note that the dynamic range that is shown for \fwciv\ and \fwmg\ coincides (1.6dex). The same situation occurs with the dynamic range that is shown for \LPcs, \LPcc\ and \LPc\ (2.5dex).   We also show the correlation coefficient r$_{\rm P}$. In all these cases P$_{\rm P}<<10^{-8}$ and are not shown in the panels.     }
 \label{fig:correlationsDR7}
 \end{figure*}

\begin{figure*}
       \includegraphics[width=1.0\textwidth ,angle =0,keepaspectratio]{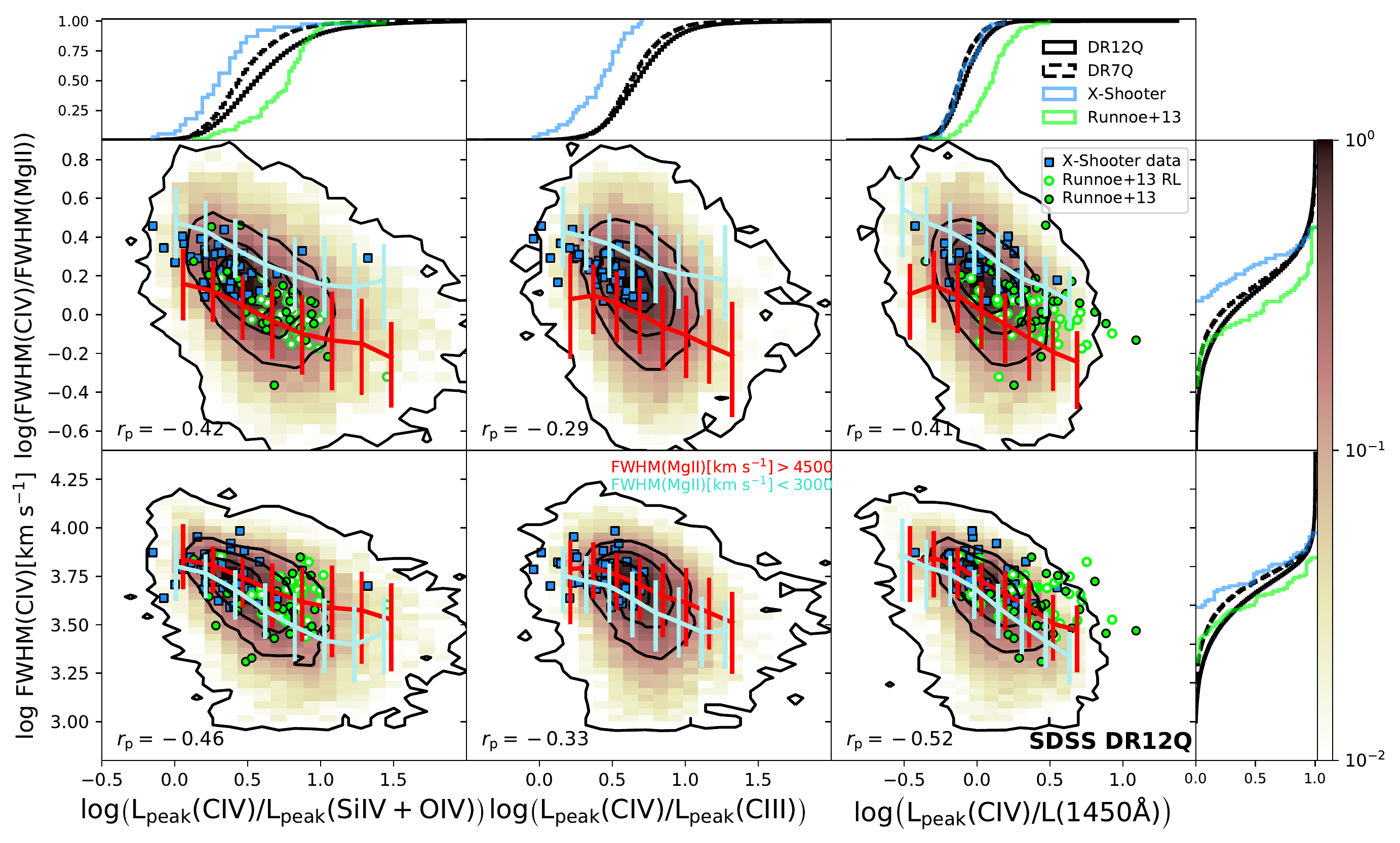}
 \caption{ As in Figure \ref{fig:correlationsDR7} but 
for the {\bf SDSS DR12Q} sample.  }
 \label{fig:correlationsDR12}
 \end{figure*}

\begin{figure*}
    \includegraphics[width=\textwidth,angle =0]{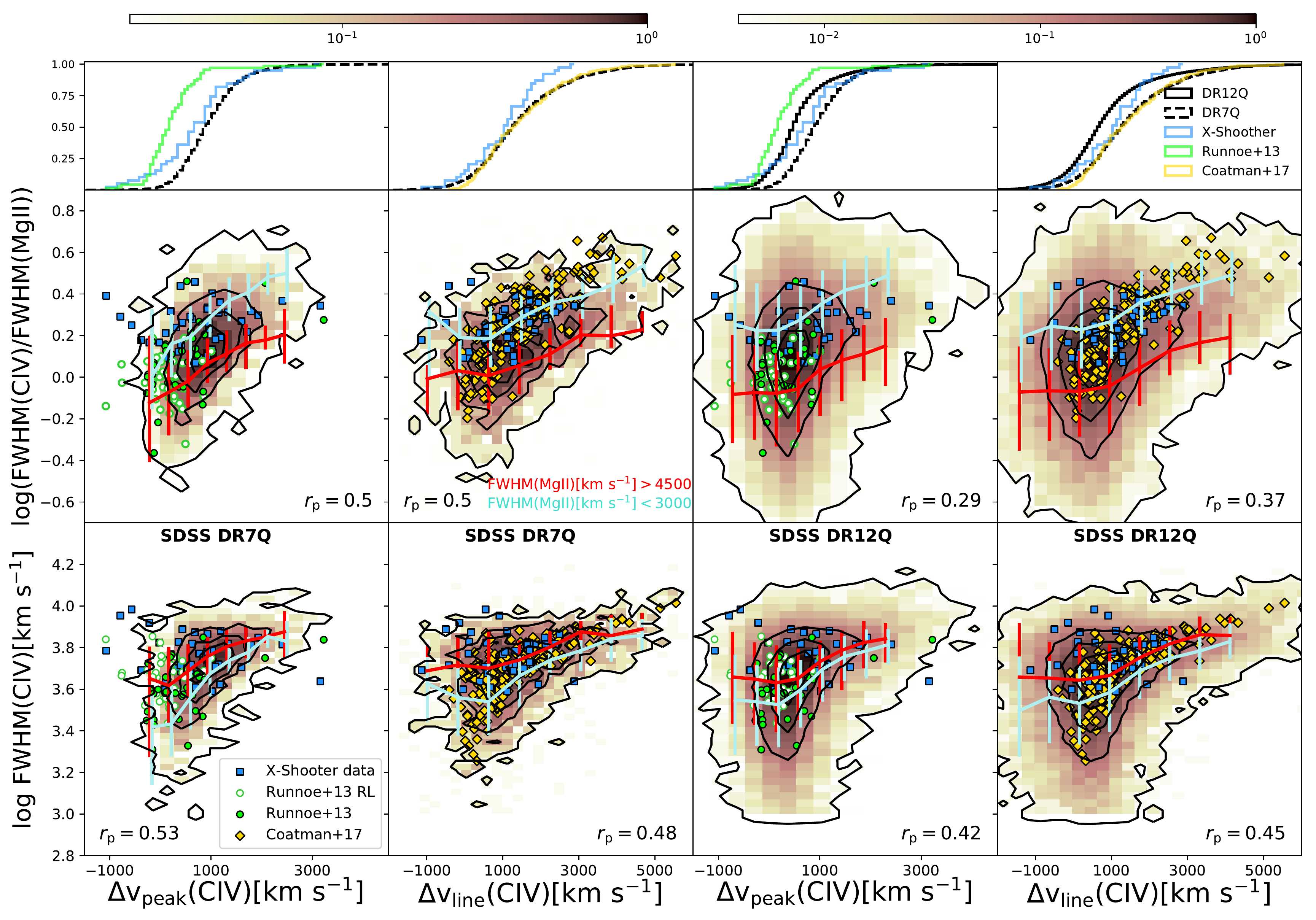}
 \caption{ Bidimensional distributions of $\fwciv/\fwmg$ (top-panels) and  $\fwciv$ (bottom panels) vs the \civ\ blueshift proxies  $\Delta v_{\rm peak}$(left) and $\Delta v_{\rm line}$ (right) in the SDSS DR7Q  (two left columns) and DR12Q (two right columns) samples. The intensity of the colour represents the relative density of points as shown in the colour bar on the top.  The black thin lines represent the   25\%, 50\%, 75\%  and 99\%   contours centred at the maximum probability point. The small samples are superimposed as indicated in the legends. The projected CDFs of $\Delta v_{\rm peak}$ and $\Delta v_{\rm line}$ are also shown in the top diagrams. In C17 sample there are not  available measurements for the \fwmg. We then used   $0.75\fwha$ as a proxy for \fwmg\ where 0.75  represent the median value of  $\fwmg/\fwha$ in the X-Shooter sample from \citet{MejiaRestrepo2016a}. Coloured trend lines represent the median values of  \fwcm\ (middle panels) and \fwciv\ (bottom panels) as a function of the different blueshift estimators  for  objects  with $\fwmg<3000\ \kms$ (light-turquoise) and $\fwmg>4500\ \kms$ (red). The error bars represent the 1-$\sigma$ dispersion of the points around these trends.  Measurements of the blue-shift are less reliable in the DR12Q sample than in the DR7Q sample because of the redshift determination  (see \S \ref{sec:data1}). We also show the correlation coefficient r$_{\rm P}$. In all these cases P$_{\rm P}<10^{-90}$ and are not shown in the panels.  }
 \label{fig:correlationsdl}
 \end{figure*}

From all the quantities considered here, the most relevant parameters for our analysis  are the following:
\begin{itemize}
\item \Luv
\item \Lthree
\item \fwciv
\item \fwmg 
\item $\fwcm \equiv \FWCM$ 
\item $\LPcs\equiv \lpcs$
\item $\LPcc\equiv \lpcc$
\item $\LPc\equiv \lpc$
\item \dvline, blue-shift of the \civ\ line centroid. 
\item \dvmax, blue-shift of the \civ\ line peak.  
\end{itemize}
In Figures \ref{fig:pdfL1450}, \ref{fig:pdfFWCIVFWMG}, \ref{fig:correlationsDR7}, \ref{fig:correlationsDR12} and \ref{fig:correlationsdl} we present relevant information associated with these quantities. First,  In Fig. \ref{fig:pdfL1450} we show 
the bi-dimensional distribution of $\log \fwciv$ versus $\log \Luv$ (left column) and $\log \fwmg$ versus $\log \Lthree$  (right column)    for the DR7Q (top panels)    and the DR12Q (bottom panels) samples. We continue with Fig. \ref{fig:pdfFWCIVFWMG} where we  show 
the bi-dimensional distribution of $\log \fwciv$ versus $\log \fwmg$   for the DR7Q (left) and the DR12Q (right) samples. We also show in Figures \ref{fig:correlationsDR7} and \ref{fig:correlationsDR12}  the bi-dimensional distributions of $\log \fwciv$  and $\log \fwcm$ versus $\log \LPcs$ (left), $\log \LPcc$ (centre) and $\log 
\LPc$ (right) for the DR7Q and the DR12Q samples, respectively.   Finally, in Figure  \ref{fig:correlationsdl}  we show  the  bi-dimensional distributions of $\log \fwcm$ versus \dvmax\ and \dvline\ for the 
DR7Q (columns 1 and 2 from left to right) and the DR12Q ( columns 3 and 4) samples, respectively. We also  show the   cumulative distribution funtion (CDF) of all the quantities and superimpose the relevant information from the X-Shooter, 
R13 and C17 samples whenever available. In all these figures we show the Pearson correlation coefficient (r$_{\rm P}$)  for the  DR7Q and DR12Q bi-dimensional distribution of 
the associated quantities.  The associated probability for upholding the null hypothesis (P$_{\rm P}$) is also shown whenever P$_{\rm P}> 1\times 10^{-8}$ otherwise it is approximated to P$_{\rm P}=0$. 

As can be seen in all these figures, all samples used for this paper are subject to different limitations and, potentially, different selection effects and biases. On the one hand, large samples have the advantage of better sampling the overall quasar population. However, they not only have limited data quality but are also incomplete at low luminosities (because of flux limits) and high luminosities  \citep[because of the upper redshift cuts;][]{Labita2009}. On the other hand, our small samples  have very good data quality but cannot statistically represent the overall quasar population. For further details, in Appendix \ref{app:samplecomp} we  discuss the particular advantages and limitations related to the large and small samples used for this work.

\section{Results and Discussion}
\label{sec:results1}

In this section we explore in detail the  different methods suggested by the aforementioned authors to improve \civ-based \Mbh\ estimations. First we will analyse our results from the largest SDSS DR7Q and DR12Q samples to subsequently contrast them with those obtained from the  X-Shooter, R13 and C17 samples and discuss the possible problems in the analysis  done with such small samples.

\subsection{SDSS DR7Q and DR12Q samples}

In Table \ref{tab:correlation} we present the correlation matrix associated with the most relevant measurements relating the \civ\ and \mgii\ lines, and the continuum emission from the accretion disk in both SDSS samples. One 
important result shown in this table, as well as in Figure \ref{fig:pdfFWCIVFWMG}, is the  very weak (or absent) correlation between \fwciv\ and \fwmg\ ($0<r_{\rm P}<0.13$, $P_{\rm P}<2\times 10^{-13}$) that inhibits the possibility  to derive reliable \civ-based \Mbh\ estimations by only comparing the \civ\  with the \mgii\ line widths. One alternative to overcome this issue is by means of a correlations between the ratio of \fwciv\ to the \fwhm\ of the low ionization lines, and other  emission lines and/or continuum property. This would provide a simple procedure to predict the width of low ionization lines in terms of \fwciv\ and other observed properties as already proposed by \citet{Runnoe2013}, \citet{MejiaRestrepo2016a} and \citet{Coatman2017}.

\subsubsection{Line Peak Ratios}
\label{sec:lpratios}

We first explore   the statistical significance of the anti-correlations that link \allowbreak \fwcm\ with \LPcs, \LPcc, and \LPc\ and which are used to improve \civ-based \Mbh\ estimations.  The reason to include  \LPc\ in this analysis,  which has not been considered in the literature,  is  its independence on other emission lines\footnote{We also considered the possibility of using the EW$\left(\civ\right)$ for our analysis. However, it shows weaker correlations with \fwcm\ and \fwciv\ than \LPcs, \LPcc, and \LPc.}. Hereafter we will refer to these three quantities as the line peak ratio quantities.

Figures  \ref{fig:correlationsDR7} and   \ref{fig:correlationsDR12} show that  \fwcm\ as well as \fwciv\ are anti-correlated with \LPc, \LPcs\ and \LPcc\ in both  SDSS quasar samples\footnote{Notice that Figures  \ref{fig:correlationsDR7} and   \ref{fig:correlationsDR12}  map the same dynamical range for $\log \fwciv$ and $\log \fwcm$ (1.6 dex in both cases) as well as for $\log \LPcs$, $\log \LPcc$ and $\log \LPc$ (a total of 2.5 dex in all of them).}. Additionally,  the corresponding values of r$_{\rm p}$ suggests that in most cases the anti-correlations of the three line peak ratio quantities with \fwciv\ are tighter than those with \fwcm\ with the exception of \LPcc\ in the DR7 sample where the anti-correlations are of comparable strength (Fig \ref{fig:correlationsDR7} middle column). In addition to this, the data presented in Table \ref{tab:scatter} shows that the scatter of the correlations associated with \fwciv\ are smaller than in those associated with \fwcm\ in both SDSS quasar samples.

\begin{table}
\caption{ Scatter found in correlations  between the listed quantities in the DR7Q and DR12Q samples }
\tabcolsep=0.25cm
\centering
\begin{tabularx}{\columnwidth}{  l l l l l  }
\hline
& \multicolumn{4}{c}{--------- scatter ----------}\\
  & \multicolumn{2}{c}{\fwcm\ } & \multicolumn{2}{c}{\fwciv\ }   \\ 
& DR7Q & DR12Q & DR7Q & DR12Q \\
\hline
          
	\LPcs & 0.21  & 0.27  &0.17 &  0.21\\ 
	\LPcc & 0.22 & 0.29  & 0.19 & 0.23\\  
	\LPc & 0.20 & 0.25 & 0.14 & 0.19 \\ 
	\dvmax & 0.19& 0.28 & 0.16 & 0.22\\ 
    \dvline & 0.21 & 0.29 & 0.19 & 0.24\\ 
\hline
\end{tabularx}
\label{tab:scatter}
\end{table}

One possibility to explain this behaviour is that the correlations related to \fwcm\   are driven by the  more fundamental \fwciv\ correlations. This interpretation is supported by   the tight correlation between \fwciv\ and \fwcm\ that we find in both  SDSS samples ($r_{\rm p}=0.71$ in both cases). Thus, \fwcm\ is just increasing the scatter of the original correlations with \fwciv. 

To test this idea we first divided our DR7Q and DR12Q samples into two sub-groups: Objects with $\fwmg<3000\ \kms$ (narrow-group) and objects with $\fwmg>4500\ \kms$ (broad-group). Then, we binned each group by the line peak ratio quantities with a bin size of 0.2 dex. For each bin we computed the median value of \fwcm\ and \fwciv\ and the corresponding 16 and 84 percentiles to quantify the dispersion in each bin.  The  light-turquoise and red solid lines in Figs. \ref{fig:correlationsDR7} and \ref{fig:correlationsDR12} represent the median values and corresponding 1-$\sigma$ dispersion in the narrow- and broad- groups, respectively.  We can see that   light-turquoise and red lines are very close to each other and that their dispersion bars overlap  in the  diagrams associated with \fwciv\ (bottom panels).  In those panels the  median red lines are just {\it slightly above} the light-turquoise lines (roughly 0.07 dex) in all diagrams. 
However, in those  diagrams associated with \fwcm\ (middle panels)  we can see a clearer separation between light-turquoise and red lines. Particularly red lines ($\fwmg>4500\ \kms$) are now roughly 0.3 dex {\it under} the light-turquoise lines ($\fwmg<3000\ \kms$) in all diagrams. This indicates that \fwmg\ is driving the dispersion  in the correlation between \fwcm\ and the line peak ratio quantities.

To obtain further support for the previous finding, we looked at the residuals of the line peak ratios when expressed as a function of \fwcm. In the case that \fwmg\ is driving the dispersion in the   \fwcm\ correlations,  we  would find significant anti-correlation between these residuals and \fwmg. To address this, we   fit the line peak ratios in terms of \fwciv\ and \fwcm\ using  bisector linear regressions. We find that for the  peak ratios as functions of    \fwcm,  all the line-peak-ratio-residuals are significantly anti-correlated with \fwmg\  ($\abs{r_{p}}>0.41$ in both samples) as expected. Moreover, for  the line peak ratios  versus \fwciv\  we do not find any significant correlations between any of the residuals with \fwmg\    ($\abs{r_{p}}<0.23$ in both samples).

An additional test consists of  estimating the statistical significance of the difference between the correlation coefficients associated with \fwciv\ and those associated with \fwcm\ in both SDSS quasar samples. The William's test,  using Fisher-z transformations,   provides a procedure to test  the relative significance of the difference between two Pearson correlation coefficients obtained from the same sample and sharing one common variable   \citep{DunnClark1969}. By applying this method to the correlations of  \fwciv\ and \fwcm\ with the common variable \LPcs, in both SDSS quasar samples we find an associated probability of  $P_{\rm William}<10^{-5}$ for upholding the null hypothesis that both correlation coefficients are equal. This result  discards  the equivalence of both  the \fwciv-\LPcs\ and the \fwcm-\LPcs\ correlations coefficients. We  find  similar behaviours for the case of the \fwciv-\LPc\ and \fwcm-\LPc\ correlations where we find    $P_{\rm William}<10^{-14}$ in both samples. Finally, for the case of the  \fwciv-\LPcc\ and \fwcm-\LPcc\ correlations we find  $P_{\rm William}=10^{-9}$ in the DR12Q sample while for the DR7 samples the correlation coefficients are identical (see Fig. \ref{fig:correlationsDR7}).

From all the evidence that we have collected, we can conclude   that  the prescriptions proposed by  \citet{Runnoe2013} and \citet{MejiaRestrepo2016a} are of limited applicability for correcting \civ-based estimates of \Mbh\ because the correlations between  the line peak ratios and  \fwciv\ are statistically stronger and  very likely driving the weaker correlations associated with \fwcm.

\subsubsection{\ion{C}{iv} blueshifts}

We  continue to test whether or not the use of \dvline\ proposed by  \citet{Coatman2017} can be used to improve \civ-based measurements. In addition to \dvline\  we will also include \dvmax\ in our analysis. The reason for this choice is the better	 stability of  \dvmax\ to continuum placement as we discuss in  \S\ \ref{sec:data1}.

In Fig.\ref{fig:correlationsdl} we show the bi-dimensional distribution of \fwcm\  and \fwciv\  versus the \civ\ blue-shift indicators \dvmax\ and \dvline\ in both SDSS samples.  In each panel we  map the same dynamic range for \fwciv\ and \fwcm. We also present the Pearson correlation coefficients for each diagram.  

Fig.\ref{fig:correlationsdl} demonstrates that  \fwcm\ and \fwciv\ are  both correlated with \dvmax\ and \dvline\  in both SDSS samples. It is also noticeable that in most cases,  the correlations between  both blue-shift estimators and \fwciv\ are tighter than with \fwcm. The only   exception is with \dvline\ in the DR7Q sample where both correlations  show similar significance. We can also notice in Table \ref{tab:scatter}  that the scatter of the  \fwciv\ correlations  is smaller than the scatter in the corresponding    \fwcm\  correlations in  both  SDSS samples. These results would indicate that the correlations associated with \fwciv\ are driving the correlations associated with \fwcm, similarly to what we found in \S \ref{sec:lpratios}.  

We repeated the same three tests described in \S \ref{sec:lpratios} to further check the reliability of these results. First, when dividing the samples into two subsets according to their \fwmg\ and binning by the blue-shift indicators in bins of 700 \kms, we find that the separation of the median trends of the narrow-group   from the median trends of the broad-group  (light-turquoise and red lines in Fig. \ref{fig:correlationsdl}) is increased from roughly -0.08 dex in the \fwciv\ diagrams to roughly 0.25 dex in the \fwcm\ diagrams. Second, the  residuals 
of the blue-shift indicators expressed as a function of \fwcm\   show significant anti-correlations with \fwmg\  ($\abs{r_{s}}>0.36$ in both samples). In contrast, when the  blue-shift indicators are expressed as a function of \fwciv\ we find  no correlations with \fwmg\    ($\abs{r_{s}}<0.10$ in both samples). Finally,  the relative significance test shows that \fwciv\ correlations are indeed stronger than the \fwcm\ correlations  (P$_{\rm William}<0.006$) for  \dvmax\ and \dvline\  in the DR12Q sample and for \dvmax\ in the DR7Q sample. In the case of \dvline\ for the DR7 sample we find that its Pearson correlation coefficients with  \fwciv\  and  \fwcm\ (0.48 and 0.50, respectively) are not statistically different  to each other  (P$_{\rm William}=0.1$, smaller than two-sigma significance). All these results support the idea that the correlations between the CIV blue-shifts and \fwciv\ are the main drivers for the correlations with \fwcm. This, in turn, suggests that the prescription introduced by   \citet{Coatman2017} may have limited applicability to improve \civ\ mass measurements.

\begin{table*}
\tabcolsep=0.06cm
\centering
\caption{Pearson correlation coefficients of  (1) $\log \fwciv$  and (2) $\log \fwcm$  versus the line peak quantities in (a)  the X-Shooter sample, (b) the R13 sample, (c) the combination of the X-Shooter and R13 samples and (d)  the combination of the X-Shooter and R13 samples excluding the RL objects. \LPcc\ measurements are not available for the R13 sample.}
\label{tab:XshR13}
\begin{tabular}{ccccccccccccccccc}

\hline      & \multicolumn{8}{c}{$\log\ \fwciv$} & \multicolumn{8}{c}{$\log\ \fwcm$}  \\ 
     & \multicolumn{2}{c}{X-Shooter$^{a}$} & \multicolumn{2}{c}{R13$^{b}$} & \multicolumn{2}{c}{Both$^{c}$} & \multicolumn{2}{c}{Both RQ$^{d}$}  & \multicolumn{2}{c}{X-Shooter$^{a}$} & \multicolumn{2}{c}{R13$^{b}$} & \multicolumn{2}{c}{Both$^{c}$} & \multicolumn{2}{c}{Both RQ$^{d}$}\\  
 & r$_{\rm P}$  &  P$_{\rm P}$          & r$_{\rm P}$  &  P$_{\rm P}$              & r$_{\rm P}$  &  P$_{\rm P}$       & r$_{\rm P}$  &  P$_{\rm P}$   &   r$_{\rm P}$  &  P$_{\rm P}$  & r$_{\rm P}$  &  P$_{\rm P}$  & r$_{\rm P}$  &  P$_{\rm P}$  & r$_{\rm P}$  &  P$_{\rm P}$      \\ \hline
$\log\ \LPcs$ & -0.21  &     2$\times 10^{-1}$         & -0.04  &  8$\times 10^{-1}$             & -0.45  &   1$\times 10^{-8}$      & -0.54 & 1$\times 10^{-6}$  &   -0.32 & 4$\times 10^{-2}$ & -0.60 & 3$\times 10^{-8}$ & -0.65 & 1$\times 10^{-11}$ &-0.54  &   1$\times 10^{-6}$     \\
$\log\ \LPc$  & -0.48 &   2$\times 10^{-3}$           & -0.30  &  1$\times 10^{-2}$              & -0.53   &   3$\times 10^{-9}$     & -0.60  & 3$\times 10^{-8}$ &     -0.40 & 1$\times 10^{-2}$ & -0.36 & 2$\times 10^{-3}$ &-0.59 & 2$\times 10^{-11}$ &-0.59 & 8$\times 10^{-8}$     \\
$\log\ \LPcc$ & -0.23 &   2$\times 10^{-1}$     &-- &     & -- &           & -- &           & -0.60 &   4$\times 10^{-5}$     & --  &       & --   &        & --  &  \\ \hline       
\end{tabular}
\end{table*}

\subsection{Small Samples}
 
 We continue our analysis exploring the small samples described in \S \ref{sec:data1}. Below we analyse the behaviour of the line peak ratios and the blue-shift  relations with \fwciv\ and \fwcm\ in those samples and discuss the similarities and differences with respect to our findings in the large SDSS samples.

\begin{table*}
\tabcolsep=0.08cm

\centering
\caption{Absolute values of the Pearson correlation coefficients ($\abs{r_{\rm p}}$)  for the quantities in the first column versus (1)\ $\log \fwciv$  and (2) $\log \fwcm$. The central values correspond to the medians  obtained  from the 100 randomly generated sub-samples  selected to have flat distributions  in   \Luv\ (a), \fwciv\ (b), \fwcm\ (c), and \Luv-\fwciv\ (d) from the SDSS DR7Q and DR12Q samples. Errors correspond to the central 68\% of the $\abs{r_{\rm p}}$ distribution. }
\renewcommand{\arraystretch}{2}
\label{tab:resampling}
\resizebox{\textwidth}{!}{\begin{tabular}{lllllllllllllllll}
\hline Random Sampling 	$\Rightarrow$ & \multicolumn{4}{c}{\Luv$^{a}$}                                                                                        & \multicolumn{4}{c}{\fwciv$^{b}$}                                                                                      & \multicolumn{4}{c}{\fwcm$^{c}$}                                                                                       & \multicolumn{4}{c}{\Luv-\fwciv$^{d}$}                                                                                 \\

                & \multicolumn{2}{c}{DR12Q}                               & \multicolumn{2}{c}{DR7Q}                                & \multicolumn{2}{c}{DR12Q}                               & \multicolumn{2}{c}{DR7Q}                                & \multicolumn{2}{c}{DR12Q}                               & \multicolumn{2}{c}{DR7Q}                                & \multicolumn{2}{c}{DR12Q}                               & \multicolumn{2}{c}{DR7Q}                                \\
                
                & \multicolumn{1}{c}{1} & \multicolumn{1}{c}{2} & \multicolumn{1}{c}{1} & \multicolumn{1}{c}{2} & \multicolumn{1}{c}{1} & \multicolumn{1}{c}{2} & \multicolumn{1}{c}{1} & \multicolumn{1}{c}{2} & \multicolumn{1}{c}{1} & \multicolumn{1}{c}{2} & \multicolumn{1}{c}{1} & \multicolumn{1}{c}{2} & \multicolumn{1}{c}{1} & \multicolumn{1}{c}{2} & \multicolumn{1}{c}{1} & \multicolumn{1}{c}{2} \\
                \hline 
\LPcs           & $0.63^{+0.03}_{-0.03}$     & $0.55^{+0.03}_{-0.04}$    & $0.59^{+0.10}_{-0.11}$     & $0.52^{+0.09}_{-0.09}$    & $0.58^{+0.05}_{-0.04}$     & $0.55^{+0.05}_{-0.05}$    & $0.61^{+0.08}_{-0.06}$     & $0.54^{+0.08}_{-0.08}$    & $0.58^{-0.06}_{-0.04}$     & $0.57^{+0.04}_{-0.05}$   & $0.62^{+0.08}_{-0.10}$     & $0.57^{+0.08}_{-0.08}$    & $0.49^{+0.02}_{-0.02}$     & $0.47^{+0.02}_{-0.02}$    & $0.61^{+0.04}_{-0.04}$     & $0.56^{+0.05}_{-0.04}$    \\
\LPcc           & $0.37^{+0.04}_{-0.04}$     & $0.37^{+0.04}_{-0.04}$    & $0.37^{+0.11}_{-0.11}$     & $0.37^{+0.11}_{-0.12}$    & $0.42^{+0.05}_{-0.06}$     & $0.41^{+0.04}_{-0.05}$    & $0.44^{+0.11}_{-0.11}$     & $0.42^{+0.11}_{-0.12}$    & $0.42^{+0.06}_{-0.04}$     & $0.37^{+0.05}_{-0.07}$   & $0.49^{+0.07}_{-0.13}$     & $0.49^{+0.09}_{-0.10}$    & $0.36^{+0.03}_{-0.02}$     & $0.36^{+0.02}_{-0.03}$    & $0.46^{+0.05}_{-0.06}$     & $0.42^{+0.05}_{-0.05}$    \\
\LPc            & $0.70^{+0.03}_{-0.02}$     & $0.56^{+0.03}_{-0.04}$    & $0.72^{+0.06}_{-0.08}$     & $0.52^{+0.09}_{-0.11}$    & $0.62^{+0.03}_{-0.04}$     & $0.52^{+0.05}_{-0.04}$    & $0.67^{+0.05}_{-0.08}$     & $0.53^{+0.09}_{-0.06}$    & $0.59^{+0.04}_{-0.05}$     & $0.49^{+0.05}_{-0.05}$   & $0.73^{+0.07}_{-0.09}$     & $0.61^{+0.09}_{-0.09}$    & $0.57^{+0.01}_{-0.02}$     & $0.49^{+0.02}_{-0.02}$    & $0.69^{+0.03}_{-0.04}$     & $0.54^{+0.05}_{-0.04}$    \\
\dvmax          & $0.48^{+0.04}_{-0.05}$     & $0.42^{+0.02}_{-0.03}$    & $0.60^{+0.09}_{-0.11}$     & $0.56^{+0.08}_{-0.09}$    & $0.44^{+0.05}_{-0.06}$     & $0.37^{+0.06}_{-0.05}$    & $0.62^{+0.08}_{-0.09}$     & $0.58^{+0.08}_{-0.10}$    & $0.42^{+0.06}_{-0.05}$     & $0.34^{+0.06}_{-0.06}$    & $0.67^{+0.06}_{-0.11}$     & $0.64^{+0.06}_{-0.06}$    & $0.48^{+0.02}_{-0.02}$     & $0.41^{+0.03}_{-0.02}$    & $0.60^{-0.05}_{+0.04}$     & $0.57^{+0.05}_{-0.04}$    \\
\dvline         & $0.62^{+0.03}_{-0.03}$     & $0.54^{+0.04}_{-0.04}$    & $0.54^{+0.08}_{-0.09}$     & $0.56^{+0.08}_{-0.09}$    & $0.42^{+0.05}_{-0.06}$     & $0.41^{+0.05}_{-0.06}$    & $0.55^{+0.09}_{-0.09}$     & $0.55^{+0.09}_{-0.08}$    & $0.42^{+0.04}_{-0.05}$     & $0.40^{+0.05}_{-0.05}$    & $0.63^{+0.07}_{-0.10}$     & $0.65^{+0.06}_{-0.08}$    & $0.46^{+0.02}_{-0.02}$     & $0.45^{+0.02}_{-0.03}$    & $0.52^{+0.04}_{-0.05}$     & $0.56^{+0.04}_{-0.05}$   \\
                \hline 
\end{tabular}}
\end{table*}

\subsubsection{Line Peak Ratios}
Given that line peak information is only available for the X-Shooter and R13 samples we limit the analysis of the line peak ratios to these two samples. In addition to the SDSS data, in Figures \ref{fig:correlationsDR7} and \ref{fig:correlationsDR12} we also show the data points and distribution functions associated  with the X-Shooter (light-blue squares) and R13 samples (lime open dots and lime filled dots for RL and RQ objects, respectively).

In Table \ref{tab:XshR13} we show the correlation coefficients of   the line peak quantities versus (1) \fwciv\  and \allowbreak (2) \fwcm\  in the following configurations:
\begin{itemize}
\item The individual X-Shooter and R13 samples (subsamples {\it a} and {\it b}).
\item The combination of the X-Shooter and the R13 samples including RL objects (subsample {\it c}).
\item The combination of the X-Shooter and the R13 samples excluding RL objects (subsample {\it d}).
\end{itemize}

 We  remark that both the X-Shooter and the R13 samples are not complete. Indeed,  as discussed in Appendix \ref{app:samplecomp}, the two samples are mapping totally different regions in the parameter spaces determined by (1) the line peak quantities versus \fwcm\  and by (2) the  line peak quantities versus  \fwciv\ (see also Figures \ref{fig:correlationsDR7} and \ref{fig:correlationsDR12}).
 Consequently, after the exclusion of the RL objects from the  R13 sample,  the  combination of the X-Shooter and the R13 samples (i.e. subsample {\it d}) maps the parameter space of RQ type-1 AGN  over wider ranges in \Luv, \fwciv, \fwcm\ and the line peak quantities.

The correlation test presented in Table \ref{tab:XshR13} suggests that for the individual samples ({\it a} and {\it b}) in most cases  the correlations of  \fwcm\ with the line peak quantities are tighter than those associated with \fwciv, in contrast to what we find for the large SDSS samples. The same behaviour is found for the combination of both samples  including the RL objects (subsample {\it c}). However, when the RL objects are excluded from the analysis (subsample {\it d}), the \fwcm\ and \fwciv\ correlations coefficients are statically  indistinguishable. Thus, the  results from subsample {\it d},  that better maps our parameter space, are in  agreement with our results from the large SDSS samples. This may indicate that  the large fraction of RL objects in the R13 sample (37/69) are probably artificially strengthening the correlations associated with \fwcm. This is probably caused by relatively broad \mgii\ profiles shown by RL objects  ($\log \fwmg[\kms] \gtrsim 3.5$, see right panel of Fig. \ref{fig:pdfL1450}).

\subsubsection{\civ\ blueshifts}

Before reporting the results of our comparative analysis for the C17 sample, we note that this sample leans towards  high-luminosity sources as can be seen in Fig. \ref{fig:pdfL1450}. Nonetheless, its   \dvline\ and \fwciv\  distribution are in very good agreement with the SDSS-DR7Q sample  (see  Appendix \ref{app:samplecomp} and Figures \ref{fig:pdfL1450}, \ref{fig:correlationsdl}  for  detalis).

Using the results reported in \citet{Coatman2017}, we find that in their sample \dvline\ is very tightly correlated with \fwciv\ ($r_{\rm p}=0.82$). However, we also  find that   the \dvline-$\fwcha$ Pearson correlation coefficient is essentially equal ($r_{\rm p}=0.83$). It is also remarkable that   Fig. 9 in \citet{Coatman2017}  shows  that the scatter in  the $\fwcha$ vs \dvline\ correlation  is clearly dominated by  \fwha.  These results support our hypothesis that the \fwciv-\dvline\ correlation is the driver of the $\fwcha$-\dvline\ correlation.

\subsection{Resampling tests}

Here we present different tests designed to check the validity of the findings presented above.  They consist of re-sampling our SDSS DR7Q and DR12Q samples in four different ways:
\begin{itemize}
\item Flat distribution in $\log \Luv$.
\item Flat distribution in $\log \fwhm$.
\item Flat distribution in $\log \fwcm$.
\item Flat distribution in both $\log  \Luv$ and $\log \fwhm$ simultaneously.
\end{itemize}
 The motivation for these tests is to check whether our findings are biased by the concentrated distribution in \Luv,  \fwciv\ and \fwcm\  and/or the known correlation between \Luv\ and \fwciv\ that we observe  in Fig. \ref{fig:pdfL1450}  (see also Appendix \ref{app:samplecomp}). To this end, we first divided our SDSS samples in bins of 0.5dex in \Luv\ starting at $\log\ \Luv = 45.0\ \ergs$ in the DR7Q sample and at $\log\ \Luv = 44.5\ \ergs$ in the DR12Q sample. For \fwciv\ and \fwcm\ we divided our sample in bins of 0.4dex.  To guarantee an equal number of objects  in each bin, we selected 23 objects from the DR7Q sample and 100 objects from the DR12Q sample, in each realization of the re-sampling simulation. We finally subdivided our SDSS samples in bi-dimensional bins of \Luv\ and \fwciv\ of 0.5 and 0.4 dex respectively. For each bin we selected 13 and 30 objects in the DR7Q and DR12Q samples. In this case the total number of sources in each re-sampling test is comparable with the size of the small X-Shooter, R13 and C17 samples. To account for statistical variance because of the  limited sampling, we repeated these procedures 100 times. We find the \fwciv\ associated correlations show  larger or equivalent statistical significance than the  \fwcm\ correlations as suggested by the William's method (See Table \ref{tab:resampling} for details) which  is  consistent to what we found for the large SDSS samples.
 
\subsection{Signal to noise analysis}

We also considered the possibility that our results may be affected by the limited quality of the SDSS spectroscopic data. 
To this end, we selected a high-quality sub-samples of the SDSS DR12Q catalogue consisting of objects with $S/N>10$ at 1700\AA\ and at 3000\AA\ with a binning of 0.75\AA/pixel, following \citet{Denney2013}. We found a total of 2230 objects meeting these criteria. 

As can be seen in Table \ref{tab:SNtab}, our analysis on this high-quality sub-sample is consistent with our central findings on the entire SDSS DR12Q sample. In particular, we find that the correlations connecting the line peak quantities and the \civ\ blueshifts with \fwciv\ are stronger than those with \fwcm. 
The only exception is with \dvline, where both correlations are of comparable strength. Similarly, in Table~\ref{tab:SNtab} we also show that when we further limit our analysis to the 483 objects with $S/N>20$, the correlations related to \fwciv\ are always stronger than those related to \fwcm.

The usage of these high-$S/N$ sub-samples also allows us to test the results of several previous studies that suggested that, in high-quality spectra, the \sigline\ of \civ\ provides more accurate  \Mbh\ estimates than the \fwhm\ of \civ. Because of the lack of \Hbeta\ measurements in our sample, we used the \fwhm\ and the \sigline of \mgii\ as proxies for the \Hbeta\ measurements. 
In contrast to the results of \citet{Denney2013}, we found  weak or even no correlations between the \sigline\ of \civ\ and, the \sigline\ and \fwhm of \mgii\ in both of the high-quality sub-samples (i.e., those with $S/N>10$ and/or $>20$; correlation coefficients in the range $0.04<r_{\rm p}<0.19$). This indicates that the \sigline\ of \civ\ cannot be reliably used to provide accurate estimates of \Mbh\ even in high quality data,  mainly because of its instability to the continuum placement.

\begin{table}
\tabcolsep=0.1cm
\centering
\caption{ Pearson correlation coefficients between the listed quantities and the three first proper vectors (PV1, PV2 and PV3) obtained from principal component analysis for the DR7Q and DR12Q correlation matrices  shown in Table \ref{tab:correlation}}
\begin{tabular}{rrrrrrr}
\hline
&\multicolumn{3}{c}{DR7} & \multicolumn{3}{c}{DR12} \\
&PV1 & PV2& PV3 & PV1 & PV2& PV3
\\
Cumulative Variance &38\% & 54\% & 67\%  & 32\% & 47\% & 61\% 
\\
\hline
$\log$ \fwciv & -0.81 &  0.27 & -0.23 & -0.75 &  0.21 &  0.25 \\
$\log$ \fwmg &  0.07 &  0.86 &  0.33 &  0.10  &  0.87 & -0.30  \\
$\log$ \fwcm & -0.72 & -0.39 & -0.43 & -0.68 & -0.45 &  0.41 \\
$\log$ \Luv & -0.48 &  0.25 & -0.17 & -0.62 &  0.22 & -0.01 \\
\dvmax &  0.73 &  0.09 &  0.09 &  0.64 & -0.04 &  0.01 \\
\dvline & 0.74 & 0.15 & 0.15 & 0.71 & 0.01 &  -0.04 \\
$\log$ \LPc &  0.86 & -0.04 & -0.37 &  0.79 &  0.02 &  0.33 \\
$\log$ \LPcs &  0.77 &  0.05 & -0.18 &  0.77 & 0.00 &  0.13 \\
$\log$ \LPcc & 0.60  & 0.25 &  -0.45 & 0.61 & 0.04 & 0.30  \\
$\log$ EW$\left(\civ\right)$ &  0.47 &  0.27 & -0.74 &  0.27 &  0.38 &  0.84 \\
$\log\ \left(L_{\rm peak}\left(\civ\right)/L\left(\civ\right)\right)$ &  0.74 & -0.40  &  0.37 &  0.45 & -0.43 & -0.66 \\
$\log \left(\fwmg/\sigma\left(\mgii\right)\right)$ & -0.13 &  0.76 &  0.31 & -0.05 &  0.76 & -0.30  \\
$\log\ \left(\Lthree/\Luv \right)$ &  0.14 & -0.41 &  0.40  & -0.13 & -0.03 & -0.19 \\
\hline

\end{tabular}
\label{tab:Eigen}
\end{table}

\subsection{Principal Component analysis}
\label{app:correlation}
We conducted a principal component analysis  on this correlation matrixes presented in Table \ref{tab:correlation}  to find different  groups of interconnected variables and obtain the amount of variance driven by each group. In Table \ref{tab:Eigen} we show the correlation coefficients between the first three proper vectors  and the quantities that define them. We can observe that the first  proper vector (PV1) is responsible for 38 and 32 percent of the variance in the DR7Q and DR12Q samples respectively. In both cases,  \fwciv\ and \LPc\ show the strongest correlations with PV1 indicating that the \fwciv-\LPc\ anti-correlation drives  PV1 and consequently a large percentage of the variance in the SDSS samples. \fwcm, \Luv, \LPcs, \LPcc, \dvmax, \dvline\ and EW$\left(\civ \right)$ and  $L_{\rm peak}\left(\civ\right)/L\left(\civ\right)$ also show important correlations with PV1. However, these correlations are basically caused by the strong relations of these quantities  with \fwciv\ and \LPc. 

The second proper vector (PV2)  is responsible for 16\% and 15\% of the variance in the DR7Q and DR12Q samples, respectively, and is    strongly correlated with \fwmg\ in both  samples. It also shows a strong correlation with $\fwmg/\sigma\left(\mgii\right)$ and a strong anti-correlation with  \fwcm, which is basically inherited from their correlations with \fwmg. Given that  by definition PV2 is linearly independent of the other proper vectors,  this result reveals that \fwmg\ is basically independent of any \civ\ related quantity and may indicate that \civ\ and \mgii\ profiles show completely independent behaviours.
 
Finally, the third proper vector (PV3) drives 13\% and 14\% of the variance of the SDSS samples, respectively, and is strongly correlated with EW$\left(\civ\right)$ and $L_{\rm peak}\left(\civ\right)/L\left(\civ\right)$ which are both strongly correlated with each other because of their dependence on  $L\left(\civ\right)$. It also shows an important correlation with \fwcm\ which is not correlated with any of these quantities. Thus, PV3 does not provide a link between \civ\ and \mgii\ properties.

\section{Summary and Conclusions}
\label{sec:conclusions} 
\civ-based \Mbh\ estimations are known to be problematic. In the past few years \citet{Runnoe2013}, \citet{MejiaRestrepo2016a} and \citet{Coatman2017} provided alternative methods attempting to improve \civ-based masses.  All these methods were based on correlations between different observables associated with the \civ\ emission and  the ratio   of \fwciv\ and the \fwhm\ of  low-ionization lines (i.e. \Halpha, \Hbeta\ and \mgii). Despite the good quality of the data used in these works,  all these methods were derived using small samples with limited coverage of the parameter space of the observables involved in each method. 

Using SDSS DR7Q and DR12Q samples (which are more representative of the quasar population) we showed that all these methods are of limited applicability to improve \civ-based \Mbh\ estimations. In fact, we find  that the aforementioned methods  depend  on correlations that are actually driven by the \fwhm\ of the \civ\ profile itself and {\emph not} by an interconnection between \fwciv\  and the  \fwhm s of the low ionization lines.  Additionally, our analysis suggests that  from all the correlations that we considered with \fwciv, those that involve \LPc\ are the  tightest ones. We also find that  other quantities considered in this work (\LPcs,  \LPcc, \dvmax\ and \dvline) are all  tightly correlated with \LPc\ (see Table. \ref{tab:correlation}).

Further support for these conclusions comes from principal component analysis 
which reveals that the first proper vector  is mostly driven by the anti-correlation between \fwciv\ and \LPc. This occurs in such a way that the relations between these quantities and  \fwcm, \Luv, EW$\left(\civ \right)$, \LPcs,  \LPcc, \dvmax\ and \dvline\ are basically driven by the \fwciv-\LPc\ anti-correlation. Notably, the second proper vector is mostly driven by \fwmg\ and shows no correlation with any \civ\ related quantity. This suggests that the properties  of the \mgii\ and \civ\ profiles are independent from each other. In other words, there is no-possibility to relate the non-virialized \civ\ emission with the virialized \mgii\ emission.

A possible explanation for this  could be associated with the  fact that the more luminous a quasar  is, the lower its EW$\left(\civ\right)$. This is the well-known \civ\  Baldwin Effect \citep{Baldwin1977,Baskin2004,BaskinLaor2005,Richards2011,Ge2016BWeff}. Both the quasar luminosity  and EW$\left(\civ\right)$  are    known to be related with the \civ\ blue-shift, the \civ\ asymmetry, and the relative strength of the X-ray emission  \citep[e.g.][]{Richards2011,Shen2016}.  Indeed, if we take \LPc\ as a proxy for the EW of \civ\  and consider the anti-correlation between \fwciv\ and  \Luv, we can conclude that the very tight anti-correlation between \fwciv\ and \LPc\ can be seen as inherited from the Baldwin Effect.

Our analysis implies that the well-characterized  \Mbh\ estimations from the low ionization lines cannot be accurately predicted from the emission line properties of the \civ\ line. Consequently, systematic infra-red spectroscopic observations of large samples of quasars are required  to guarantee the coverage of low ionization lines and the proper determination of the SMBH masses at $z\gtrsim 2$.  Achieving accurate \civ\ mass estimations requires, apart from  a robust determination of the $R_{\rm BLR}\left(\civ \right)$-$\Luv$ relation, an extensive  analysis of the \civ\ emission line itself to further understand  the virialized  component of the  \civ\ line.

\section*{Acknowledgments} 

JM acknowledges ``CONICYT-PCHA/doctorado nacional para extranjeros/2013-63130316'' for their PhD scholarship and Universidad de Chile grant "Ayudas para estad\'ias cortas de investigaci\'on destinadas a estudiantes de doctorado y magister"  for their financial support to visit ETH-Z{\"u}rich, where most of this work was done. JM also acknowledges the financial support provided by the ETH-Z{\"u}rich during his stay. PL  acknowledges support by Fondecyt Project \#1161184 and H.N acknowledges support by the Israel Science Foundation grant 234/13.

\bibliographystyle{mnras}

\appendix

\section{Fitting Procedure, measurements and BALQSO exclusion}
\label{app:fit}

Our broad emission line modelling  follows the procedure presented in \citet{MejiaRestrepo2016a} and \citet{TrakhtenbrotNetzer2012}. Very briefly,  the most prominent lines (\sioiv, \civ, \ciii\ and \mgii)  are modelled using two Gaussian components while other  weak emission lines  are modelled with a single Gaussian (including He\,\textsc{ii}1640, N\,\textsc{iv}1718, Si\,\textsc{iii}]1892). The central wavelength of each Gaussian component  is restricted to move within 1000 \kms\ around the laboratory central wavelength. The \civ\ and He\,\textsc{ii}1640 are allowed to be blue shifted up to 5000 \kms. 

 \begin{table}
\tabcolsep=0.5cm
\centering
\caption{ Spectral pseudo-continuum windows used for our line fitting procedure under the \local continuum approach. $^{1}$For each object, we manually adjusted the continuum bands, using the listed wavelength ranges as a reference. }
\begin{tabularx}{0.47\textwidth}{  l l l  }
\hline
	Line Complex  & \multicolumn{2}{c}{--------- Continuum windows$^{1}$ ----------}   \\ 
  \hline
	\sioiv & 1340-1360\AA & 1420-1460\AA  \\ 
	\civ & 1430-1460\AA & 1680-1720\AA \\  
	\ciii & 1680-1720\AA & 1960-2020\AA \\ 
	\mgii & 2650-2670\AA & 3020-3040\AA \\ 
\hline
\end{tabularx}

\label{tab:cwindows}
\end{table}

We automatized the procedure by introducing some additional steps to the line and continuum fitting. We first proceed to fit and subtract the continuum emission within a pair of continuum windows around each line. These continuum windows are set at the wavelengths that we list in Table \ref{tab:cwindows}.  We subsequently fit the emission line following the ``local" approach described in  \citet{MejiaRestrepo2016a}. After this we obtain the residuals of the fitting and  remove the pixels with the 3\% most negative fluxes within the continuum windows. The purpose of this step is to exclude from the fitting strong absorption features. We repeat the entire procedure three times to guarantee convergence.  We  also exclude from our sample objects with final reduced $\chi^2$ larger than 3.

To avoid \civ\ BALQSOs  we  excluded from the sample objects with more than 7\% of the pixels with negative \civ\ residuals. To test the performance of this automatic selection method we compare the objects that are flagged as BALQSOs using our method with the sample of 562 manually classified BALQSOs  from the SDSS-DR2 quasar database that is described in \citet{Ganguly2007}. With our criterion we flagged as BALQSO a total of 573/5088 objects from the SDSS-DR2 quasar catalogue at $1.7<z<2.0$. From these objects we found that  560/562 objects are also classified as BALQSO in the manually selected \citet{Ganguly2007} sample. These results translate into a successful identification rate of 99.6\% and a false positive identification rate of 2.3\%. After excluding the BALQSO candidates and objects with unreliable fits we end up with  3267 from the DR7Q catalogue (out of the originally selected 4817 objects)  and with 35674 objects from the DR12Q catalogue (out of the originally selected 69092 objects).

\begin{table}
\tabcolsep=0.1cm
\centering
\caption{Pearson correlation coefficients  for the quantities in the first column versus (1)\ $\log \fwciv$  and (2) $\log \fwcm$. In all cases it yields P$_{P}<1\times 10^{-13}$. }
\label{tab:SNtab}
\begin{tabular}{lllllll}
\hline
             & \multicolumn{2}{c}{All objects} & \multicolumn{2}{c}{S/N\textgreater10} & \multicolumn{2}{c}{S/N\textgreater20} \\
             & \multicolumn{1}{c}{1}    & \multicolumn{1}{c}{2}  & \multicolumn{1}{c}{1}     & \multicolumn{1}{c}{2}    & \multicolumn{1}{c}{1}      & \multicolumn{1}{c}{2}    \\ \hline
$\log$ \LPc  &-0.53           & -0.41         & -0.60              & -0.48            & -0.62              & -0.47            \\
$\log$ \LPcs &-0.46           & -0.42         & -0.52              & -0.48            & -0.59              & -0.49            \\
$\log$ \LPcc &-0.33           & -0.29         & -0.34              & -0.33            & -0.41              & -0.35            \\
\dvmax       &\ 0.37            &\ 0.29          &\ 0.55               &\ 0.46             &\ 0.62               &\ 0.45             \\
\dvline      &\ 0.39            &\ 0.37          &\ 0.54               &\ 0.56             &\ 0.57               &\ 0.55     \\   \hline    
\end{tabular}
\end{table}

\begin{table*}
\tabcolsep=0.2cm
\centering
\caption{ Pearson correlation coefficients between the listed quantities for the DR7Q and DR12Q samples. We note that in both samples, whenever $\abs{\rm r_{P}}>0.1$ it yields P$_{P}<1\times 10^{-10}$.}
\begin{tabular}{rrrrrrrrrrrrrr}
\hline
& \multicolumn{12}{c}{DR7Q correlations}\\

Property  &  1   &  2 &  3 & 4 & 5 &  6 & 7 & 8 &  9 & 10 & 11  & 12 & 13 \\
\hline
$^{1}$ $\log\ \fwciv$ &  1    &  0.17 &  0.71 &  0.34 & 0.51 &  0.44 & -0.63 & -0.51 &  -0.36 & -0.11 & -0.83 &  0.22 & -0.19 \\
$^{2}$ $\log\ \fwmg$ &  0.17  &  1    & -0.58 &  0.10  &  -0.11 & -0.19 & -0.07 &  0.03 & 0.09 &  0.07 & -0.20  &  0.60  & -0.10  \\
$^{3}$ $\log\ \fwcm$ &  0.71  & -0.58 &  1    &  0.21 & 0.50  &  0.50  & -0.47 & -0.45 &  -0.36 & -0.15 & -0.55 & -0.24 & -0.09 \\
$^{4}$ $\log \Luv$ &  0.34   &  0.10  &  0.21 &  1    & 0.28 &  0.37 & -0.36 & -0.36 &  -0.07 & -0.09 & -0.44 &  0.08 & -0.16 \\
$^{5}$\dvmax & 0.51 &  -0.11 &  0.50  & 0.28  &  1    & 0.68 &  -0.49 & -0.50  &  -0.39 &  -0.25 &  -0.45 & 0.05 &  -0.07 \\
$^{6}$\dvline &  0.44& -0.19 &  0.50  &  0.37 & 0.68 &  1    & -0.51 & -0.53 &  -0.34 & -0.25 & -0.48 & -0.03 & -0.09 \\
$^{7}$ $\log\ \LPc$    & -0.63& -0.07 & -0.47 & -0.36 &  -0.49 & -0.51 &  1    &  0.65 & 0.60  &  0.77 &  0.58 & -0.20  &  0.07 \\
$^{8}$ $\log\ \LPcs$ & -0.51  &  0.03 & -0.45 & -0.36 &  -0.50  & -0.53 &  0.65 &  1    & 0.47 &  0.45 &  0.43 & -0.08 &  0.03 \\
$^{9}$ $\log\ \LPcc$ &  -0.36 & 0.09 &  -0.36 &  -0.07 & -0.39 &  -0.34 & 0.60  & 0.47 &  1    & -0.55 & 0.23 &  -0.03 &  -0.15 \\
$^{10}$ $\log\ {\rm EW} \left(\civ\right)$ & -0.11 &  0.07 & -0.15 & -0.09 &  -0.25 & -0.25 &  0.77 &  0.45 & -0.55 &  1    & -0.08 & -0.06 & -0.13 \\
$^{11}\log\ \left(L_{\rm peak}\left(\civ\right)/L\left(\civ\right)\right)$ & -0.83 & -0.20  & -0.55 & -0.44 &  -0.45 & -0.48 &  0.58 &  0.43 & 0.23 & -0.08 &  1    & -0.23 &  0.28 \\
$^{12}\log\ \left(\fwmg/\sigma\left(\mgii\right)\right)$&   0.22 &  0.60  & -0.24 &  0.08 & 0.05 & -0.03 & -0.20  & -0.08 &  -0.03 & -0.06 & -0.23 &  1    & -0.19 \\
$^{13}\log\ \left(\Lthree/\Luv\right)$ & -0.19 & -0.10  & -0.09 & -0.16 &  0.07 & -0.09 &  -0.07 &  0.03 &  -0.15 & -0.13 &  0.28 & -0.19 &  1    \\

\hline
& \multicolumn{12}{c}{DR12Q correlations}\\

\hline
$^{1}$ $\log$\fwciv &  1    &  0.14 &  0.71 &  0.39 & 0.37 &  0.39 & -0.53 & -0.46 &  -0.33 & -0.01 & -0.50  &    0.16 & -0.01 \\
$^{2}$ $\log$ \fwmg &  0.14 &  1    & -0.60  &  0.08 & 0.01 & -0.08 & -0.01 &  0.07 & 0.04 &  0.08 & -0.11 &  0.58 &  0.02 \\
$^{3}$ $\log$ \fwcm &  0.71 & -0.60  &  1    &  0.25 & 0.29 &  0.37 & -0.41 & -0.42 &  -0.29 & -0.07 & -0.32 & -0.29 & -0.03 \\
$^{4}$ $\log$ \Luv &  0.39 &  0.08 &  0.25 &  1    & 0.31 &  0.43 & -0.37 & -0.53 &  -0.27 & -0.06 & -0.29 &  0.11 &  0.14 \\
$^{5}$\dvmax & 0.37 & 0.01 & 0.29 & 0.31 &  1    & 0.66 &  -0.39 &  -0.37 & -0.29 &  -0.13 &  -0.22 & 0.01 & 0.08 \\
$^{6}$\dvline &  0.39 & -0.08 &  0.37 &  0.43 & 0.66 &  1    & -0.45 & -0.46 &  -0.29 & -0.14 & -0.27 & -0.02 &  0.03 \\
$^{7}$ $\log$ \LPc & -0.53 & -0.01 & -0.41 & -0.37 &  -0.39 & -0.45 &  1    &  0.59 & 0.53 &  0.58 &  0.28 & -0.11 & -0.04 \\
$^{8}$ $\log$ \LPcs & -0.46 &  0.07 & -0.42 & -0.53 &  -0.37 & -0.46 &  0.59 &  1    & 0.47 &  0.27 &  0.26 & -0.08 & -0.09 \\
$^{9}$ $\log$ \LPcc &  -0.33 & 0.04 &  -0.29 &  -0.27 & -0.29 &  -0.29 & 0.53 & 0.47 &  1    & -0.30  & 0.16 &  -0.07 &  -0.28 \\
$^{10}$ $\log$ EW$\left(\civ\right)$ & -0.01 &  0.08 & -0.07 & -0.06 &  -0.13 & -0.14 &  0.58 &  0.27 & -0.30  &  1    & -0.61 & 0.00    & -0.04 \\
$^{11}\log\ \left(L_{\rm peak}\left(\civ\right)/L\left(\civ\right)\right)$ & -0.50  & -0.11 & -0.32 & -0.29 &  -0.22 & -0.27 &  0.28 &  0.26 & 0.16 & -0.61 &  1    & -0.10  &  0.01 \\
$^{12} \log\ \left(\fwmg/\sigma\left(\mgii\right)\right)$&  0.16 &  0.58 & -0.29 &  0.11 & 0.01 & -0.02 & -0.11 & -0.08 &  -0.07 & 0.00    & -0.10  &  1    & -0.07 \\
$^{13} \log \left(\Lthree/\Luv\right)$& -0.01 &  0.02 & -0.03 &  0.14 & 0.08 &  0.03 & -0.04 & -0.09 &  -0.28 & -0.04 &  0.01 & -0.07 &  1    \\
\hline

\end{tabular}
\label{tab:correlation}
\end{table*}

\section{Sample comparison}
\label{app:samplecomp}
 Here we describe in detail the parameter space of the most relevant physical quantities derived for the samples and summarize the most relevant issues associated with each of the of the small and large samples used in this paper.

\subsection{Comments on large samples}

\begin{itemize}

\item In both SDSS samples we can observe that  \fwciv\ and \Luv\  correlate with each other ($r_{\rm p} \equiv r_{\rm Pearson}\sim 0.35$). This Indicates that, on average, more luminous quasars typically show  broader \civ\ line profiles. However, the \fwmg\ is completely independent of the quasar luminosity ($r_{\rm p} < 0.05$ in both SDSS samples).

\item The DR12Q sample probes considerably fainter sources, compared with the DR7Q sample (by $\sim0.5$dex), and extends to $\Luv\gtrsim 10^{44.5}\ergs$. This difference between both samples allows us to directly test the impact of luminosity limited  samples as well as data quality in our analysis.

\item  In the case of  $\log \fwciv$ and $\log \fwmg$ both  SDSS samples span  similar ranges, from  $\sim$3 to $\sim$4.2 in $\log \fwciv$, from $\sim$3 to $\sim$3.8 in $\log \fwmg$ .  However, the DR12Q sample  has a larger fraction of object with low $\log \fwciv$ and $\log \fwmg$. Explicitly, 5\% (10\%) of the objects in the DR12Q sample have  $\fwciv\lesssim 3.3$ ($\fwmg\lesssim 3.3$) versus 2\% (6\%) in the DR7Q sample. Additionally, the DR12Q sample  has a larger fraction of objects with large $\log \fwmg$.  The sharp cut at $\log\ \fwciv=\log\ \fwmg=3$ is imposed by the fitting criterion. 

\item The  $\log \fwcm$ distribution in both samples span over similar ranges, from $\sim$-0.6 to $\sim$0.8. However, the DR12Q sample shows a larger fraction (28\%) of objects with $\log \fwcm<0$  than the DR7Q sample (20\%).

\item We can observe that the DR12Q sample shows a larger fraction (roughly $10\%$) of objects with  $\log\ \LPcs$ and $\log\ \LPcc$ $\gtrsim 1$   than the DR7Q sample (roughly $\sim 5\%$). We need however to be cautious about the reliability of such measurements because those objects show $L_{\rm peak}\left(\sioiv\right)$ and $L_{\rm peak}\left(\ciii\right)$ weaker than one tenth of $L_{\rm peak}\left(\civ\right)$. Thus, the signal to noise of the \ciii\ and \sioiv\ line profiles is probably very low in many of those objects.  We can also appreciate that the DR12Q and DR7Q samples show a similar distribution in  $\log\ \LPc$.

\item As we already explained,  DR7Q blue-shift  estimations are more accurate that in DR12Q. Additionally, we can observe in Fig. \ref{fig:correlationsdl} that the \dvmax\ and the \dvline\  distributions show a larger fraction of objects with small blue-shifts in the DR12Q sample than in the DR7Q sample (6\% in DR7Q  vs 23\% DR12Q and 12\% in DR7Q vs 28\% in DR12Q for \dvmax\ and \dvline\ $<200\ \kms$ respectively) . This behaviour   is probably caused by objects in the DR12Q sample whose cosmological redshift has been estimated using the \civ\ profile. This effect  artificially biases the  \civ\ blue-shifts towards values close to 0.  Because of  these problems with the DR12Q redshift determinations we will mainly focus our blue-shift analysis on the DR7Q sample. 

\end{itemize}

\subsection{Comments on small samples}

\begin{itemize}

\item The R13 sample is mostly described by uniform distributions in \Luv, \fwciv\ and \fwmg\ that  are   fairly spread around the SDSS data. However, RL AGN are mostly high luminosity and objects and show large \fwmg\ values. We can also observe that the R13 \fwcm\ distribution is clearly shifted towards larger values than the peak of the SDSS distributions. In terms of \dvmax, the sample is  shifted  towards low values (75\% with $<1000\ \kms$). Finally, in terms of $\log \lpcs$ and $\log \lpcc$, the sample is shifted towards large values (75\% with $\gtrsim 0.5$).

\item The X-Shooter Sample also shows mostly flat distributions in $\log \Luv$, $\log \fwciv$ and $\log \fwmg$  that are also fairly spread around the SDSS data.  Its $\log \fwcm$  is clearly distributed towards  values lower  than the peak of the SDSS distribution. Line peak ratios ($\sim $75\% with  $\log \lpcs\lesssim 0.5$ and $\log \lpcc \lesssim 0.5$) and blue-shifts (75\% with $<1000\ \kms$) are both distributed towards low values with respect to the SDSS distribution peaks. 

\item The C17 sample is clearly dominated by objects with very large \Luv\ ($\Luv \gtrsim 10^{46}\ergs$) compared to the other samples. However, its \fwciv\ and \dvline\ distributions  very closely follow the SDSS DR7Q distributions. 
\end{itemize}

\end{document}